\pdfoutput=1

\documentclass[11pt]{article}

\usepackage[final]{acl}
\usepackage{float}

\usepackage{times}
\usepackage{latexsym}
\usepackage{arydshln}
\usepackage{bm}
\usepackage{makecell}
\usepackage{colortbl}
\usepackage[T1]{fontenc}

\usepackage[utf8]{inputenc}

\usepackage{microtype}

\usepackage{inconsolata}

\usepackage{kotex}
\usepackage{booktabs}
\usepackage{colortbl}
\usepackage{multirow}
\usepackage{pifont}  
\usepackage{mathrsfs}
\definecolor{lightyellow}{RGB}{255,255,224}
\newcommand{\cmark}{\textcolor{green!60!black}{\ding{51}}} 
\newcommand{\xmark}{\textcolor{red!70!black}{\ding{55}}}   

\usepackage{tcolorbox}
\usepackage{fancyvrb}
\tcbuselibrary{skins,breakable}
\usepackage{listings}

\usepackage{graphicx}
\usepackage{fvextra}

\title{Does Your Voice Assistant Remember? Analyzing Conversational Context Recall and Utilization in Voice Interaction Models}

\author{
    {\bf Heeseung Kim$^{1}$\thanks{Equal Contribution}} \hspace{6mm}  
    {\bf Che Hyun Lee$^{1}$\footnotemark[1]} \hspace{6mm}  
    {\bf Sangkwon Park$^{1}$} \hspace{6mm}
    {\bf Jiheum Yeom$^{1}$} \\
    {\bf Nohil Park$^{1}$} \hspace{6mm}  
    {\bf Sangwon Yu$^{1}$} \hspace{6mm}  
    {\bf Sungroh Yoon$^{1,2}$\Thanks{ Corresponding Author}} \\ \\
   $^{1}$Department of Electrical and Computer Engineering, Seoul National University \\
   $^{2}$AIIS, ASRI, INMC, ISRC, and IPAI, Seoul National University \\ \\
   {\tt \fontsize{10}{10}\selectfont \{gmltmd789, saga1214, tkdrnjs0621, quilava1234, pnoil2588, dbtkddnjs96, sryoon\}@snu.ac.kr}}

\begin{document}
\maketitle
\begin{abstract}
Recent advancements in multi-turn voice interaction models have improved user-model communication. However, while closed-source models effectively retain and recall past utterances, whether open-source models share this ability remains unexplored. To fill this gap, we systematically evaluate how well open-source interaction models utilize past utterances using ContextDialog, a benchmark we proposed for this purpose. Our findings show that speech-based models have more difficulty than text-based ones, especially when recalling information conveyed in speech, and even with retrieval-augmented generation, models still struggle with questions about past utterances. These insights highlight key limitations in open-source models and suggest ways to improve memory retention and retrieval robustness.\footnote{ContextDialog: \href{https://huggingface.co/datasets/ContextDialog/ContextDialog}{https://huggingface.co/datasets/Context\\Dialog/ContextDialog}} 
\end{abstract}

\section{Introduction}

Voice assistants such as Apple Siri and Amazon Alexa have become an irreplaceable element of daily life, enabling natural and efficient speech-based interactions. In early systems, a cascaded pipeline is employed where speech is first transcribed using automatic speech recognition (ASR), then processed as text, and finally converted back to speech via text-to-speech (TTS) \cite{lin-etal-2024-advancing}. With the advent of large language models (LLMs) \cite{NEURIPS2020_1457c0d6, grattafiori2024llama3herdmodels}, however, the research community has shifted greatly towards end-to-end approaches. These models integrate ASR, text processing, and TTS into a unified multimodal framework \cite{zhang-etal-2023-speechgpt}, which not only reduces latency \cite{xie2024miniomnilanguagemodelshear} but also better preserves the richness of vocal cues \cite{kim2024paralinguisticsaware}. In line with this trend, GPT-4o \cite{openai2024gpt4ocard} has demonstrated impressive capabilities by processing visual, speech, and text data in an end-to-end manner, where various voice interaction models, datasets, and benchmarks have rapidly emerged alongside. \cite{cheng2025omnichatenhancingspokendialogue, cheng2025voxdialogue, fang2025llamaomni, xie2024miniomni2opensourcegpt4ovision}.

Despite these advances, most current models excel only in single-turn interactions. In practical applications, however, users engage in multi-turn dialogs where a one-off response is insufficient. Specifically, models must continuously retain and leverage contextual information from previous turns. For example, Gemini 2.0 \cite{google2024gemini} demonstrates the ability to remember preceding details—for instance when a user provides an apartment door code during interaction and inquires about it later—thereby showcasing robust context-maintenance. Notably, other closed-source solutions, such as OpenAI’s Advanced Voice Mode \cite{openai2024twelve}, have also showcased similar capabilities by referencing past interactions.

In parallel, the open-source community has also intensified its efforts to develop voice interaction models that support multi-round communications \cite{kyutai2024moshi, yao2024minicpm, zeng2024glm4}. Typically, these models take speech as input and generate both text and speech outputs, rather than producing spoken responses alone, to leverage the strengths of pre-trained LLMs and ensure coherent, multi-turn responses. However, it remains unclear whether current open-source systems can effectively retain and utilize long-range interaction histories. Furthermore, there are no benchmarks that explicitly require leveraging dialog history to generate responses.

In this work, we systematically investigate the ability of open-source voice interaction models to maintain and utilize conversational context through two key experiments. We evaluate (1) whether models can recall and generate spoken responses based on previous dialog and (2) their robustness in incorporating externally retrieved utterances. To support this evaluation, we introduce ContextDialog—a speech-to-speech benchmark that focus on assessing recall via spoken question-answer (QA) pairs derived from existing spoken dialogs, prompting one speaker to reference earlier information.

Our findings reveal that open-source models struggle with past speech in two key aspects. \textbf{Performance gap with text-based systems} – Speech models generally perform worse than their text-based counterparts, and \textbf{Modality-based recall gap} – Within speech models, recalling speech-based information is less accurate than retrieving text, likely due to weaker speech processing capabilities.
Additionally, our investigation of retrieval-augmented generation (RAG) shows that it fails to compensate for the model's inability to recall past information. We identify a major challenge: \textbf{Sensitivity to retrieval errors} – Models are highly susceptible to retrieval mistakes, leading to unchanged or even degraded performance.
Through these findings, we highlight the challenges models face in processing past conversational context and their sensitivity to noise in retrieved information, drawing attention to a fundamental, yet often overlooked, capability within the open-source community. Our contributions are as follows: 

\begin{itemize} 
    \item 
     We introduce ContextDialog, a benchmark designed to evaluate the models' ability to utilize dialog history in multi-turn conversations.\footnote{Project Page: \href{https://contextdialog.github.io/}{https://contextdialog.github.io/}}
    \item We show that most open-source models struggle with recalling past dialogs and fail to effectively incorporate retrieved information, even when augmented with external retriever.
    \item Through extensive evaluation and analysis, we uncover overlooked limitations in current models that restrict their applicability and propose directions for future improvements.
\end{itemize}

\section{Related Works}
\textbf{Voice Interaction Models} 
Early voice interaction models follow a cascaded approach \cite{lin-etal-2024-advancing}, transcribing speech into text, processing the transcription, and then synthesizing the output speech. Recently, end-to-end pipelines have emerged, performing these steps within a single model \cite{zhang-etal-2023-speechgpt}. Although some models generate spoken responses without relying on text \cite{nguyen-etal-2023-generative}, the inherent length and data scarcity of speech hinder semantic  modeling \cite{kyutai2024moshi}. Recent approaches integrate text generation within speech modeling to mitigate such problem, leveraging pre-trained LLMs by incorporating text as an intermediate representation \cite{kim2024paralinguisticsaware, zhang-etal-2023-speechgpt}, generating it alongside speech \cite{fang2025llamaomni}, or interleaving it with speech tokens \cite{zeng2024glm4}.

Many prior works focus on single-turn voice interaction~\cite{fang2025llamaomni, kim2024paralinguisticsaware, xie2024miniomnilanguagemodelshear, xie2024miniomni2opensourcegpt4ovision, zeng2025scaling, zhang-etal-2023-speechgpt, zhao2024advancingspeechlanguagemodels}. As a natural extension, multi-turn voice interaction models have also emerged~\cite{chen2024slamomnitimbrecontrollablevoiceinteraction, chen2025minmomultimodallargelanguage, kyutai2024moshi, fu2025vita, li2025baichuanomni15technicalreport, mai2025realtimetextlessdialoguegeneration, mitsui-etal-2024-pslm, park-etal-2024-lets, veluri-etal-2024-beyond, xiong2024freeze, yao2024minicpm, zeng2024glm4, zhang2024intrinsicvoiceempoweringllmsintrinsic, zhang2025omniflattenendtoendgptmodel, zhong2024lyra}.

Among these, only SLAM-Omni~\cite{chen2024slamomnitimbrecontrollablevoiceinteraction} and Lyra~\cite{zhong2024lyra} explicitly discuss long-context modeling. SLAM-Omni seeks to improve multi-turn modeling by storing dialog history in transcribed text form, which is then prepended as a prefix during inference. Lyra, on the other hand, explores and proposes various techniques to handle long audio histories and extends context windows directly within its model architecture. The remaining works either mention the use of multi-turn data for training or imply it through demonstrations or official implementations. However, they generally do not propose explicit methods or evaluations specifically targeting long-context recall capability. Consequently, whether these models can effectively handle past conversational history in real-world multi-turn dialog scenarios remains largely unexplored.

For voice assistants to function effectively, it is essential to assess their ability to retain and utilize prior utterances to generate contextually appropriate responses. To this end, we select SLAM-Omni (included in Appendix \ref{subsubapp:additional_models} due to its smaller model size) and Lyra as our primary long-context baselines. Additionally, we include Freeze-Omni, GLM-4-Voice, MiniCPM-o, and Moshi (also in Appendix \ref{subsubapp:additional_models}) as strong open-source candidates with multi-turn capabilities available.

\begin{figure*}
    \centering
    \includegraphics[width=0.98\linewidth]{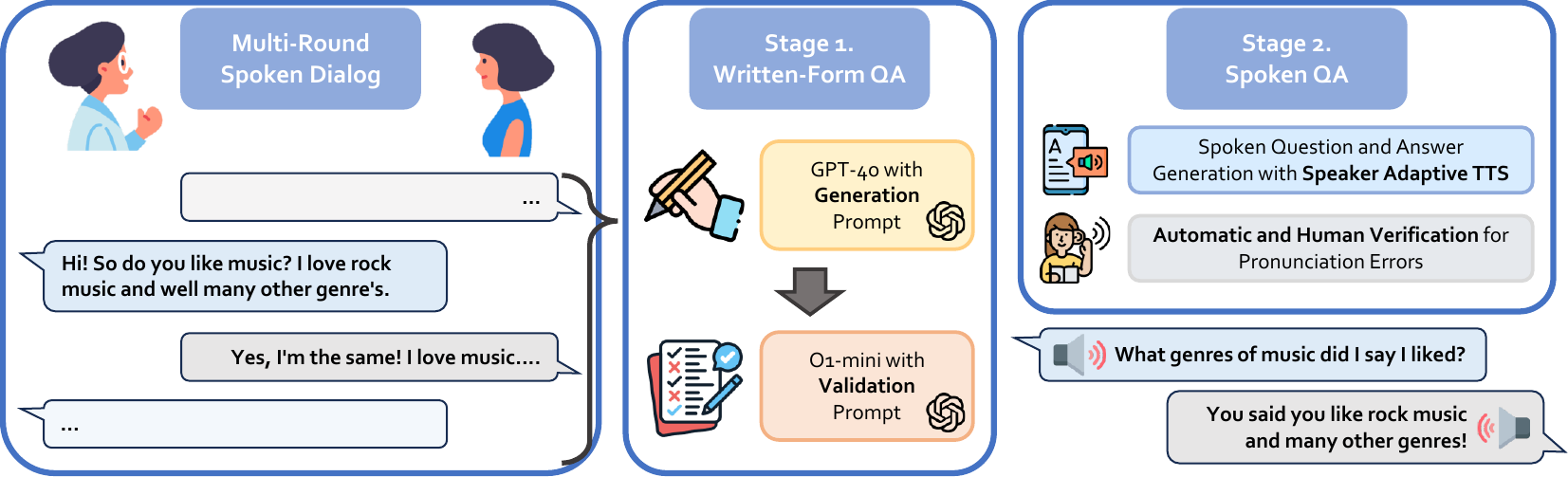}    
    \caption{Overview of the ContextDialog generation process. Past-recall QA pairs are first generated and validated (Section \ref{subsec:generation}), then converted to speech via adaptive TTS and verified both automatically and manually (Section \ref{subsec:tts}).}
    \label{fig_contextdialog}
    \vskip -0.2in
\end{figure*}

\textbf{Benchmarks} Numerous datasets and benchmarks for audio foundation models have emerged \cite{sakshi2025mmau, wang2024audiobench, yang-etal-2024-air}, particularly for voice interaction models \cite{chen2024emovaempoweringlanguagemodels, fang2025llamaomni, park-etal-2024-lets, xie2024miniomnilanguagemodelshear}. For example, in task-oriented spoken dialogs, benchmarks assess a model’s ability to recognize entities and dialog states from past utterances \cite{henderson-etal-2014-second, NEURIPS2023_7b16688a, spithourakis-etal-2022-evi}, while in open-domain dialogs, they focus on modeling and evaluating response coherence \cite{Busso2008, cieri-etal-2004-fisher, cheng2025omnichatenhancingspokendialogue, park-etal-2024-lets}. Beyond semantic relevance, recent works propose benchmarks targeting non-verbal components crucial for voice interaction models, such as gender, emotion, and background noise \cite{NEURIPS2024_681fe4ec, chen2024voicebenchbenchmarkingllmbasedvoice, cheng2025voxdialogue}. Unlike existing benchmarks that evaluate multi-turn semantics without ensuring past information is necessary for responses, ContextDialog explicitly requires models to retrieve and utilize relevant past utterances, enabling a systematic assessment of recall ability.

\textbf{Retrieval in Voice Interaction Model} 
With advancements in RAG techniques in natural language processing (NLP) \cite{JMLR:v24:23-0037, NEURIPS2020_6b493230}, efforts to integrate RAG into spoken dialog models have emerged \cite{10448210, Min2025, 10447448}. Prior works have primarily focused on task-oriented dialog for entity extraction \cite{10447448} or spoken question answering \cite{10448210, Min2025}, retrieving information from long speech documents \cite{lee2018spoken}. In contrast, we focus on multi-turn voice interactions, examining whether relevant data retrieved via an external module can be effectively utilized in the generation, specifically tailored for recent open-source interaction models.

\section{ContextDialog}
\label{sec:dataset}

We propose ContextDialog, a comprehensive benchmark designed to evaluate a voice interaction model’s ability to engage in, retain, and leverage relevant information throughout multi-turn conversations, reflecting real-world scenarios where people often forget and revisit past exchanges. ContextDialog is constructed using MultiDialog \cite{park-etal-2024-lets}, a spoken dialog corpus featuring conversations between two speakers, comprising approximately 340 hours of data with at least 10 turns per conversation from 12 speakers. 
We use the \texttt{test\_freq} and \texttt{test\_rare} splits from MultiDialog, consisting of 450 and 381 spoken dialogs, respectively. Some data are filtered during generation and validation, with the final statistics of ContextDialog shown in Table \ref{tab:contextdialog_stats} and the data generation pipeline illustrated in Figure \ref{fig_contextdialog}.

\begin{table}[t]
    \centering
    \small
    \begin{tabular}{l|c|cc}
        \toprule
        \textbf{} & \textbf{Statistics} & \textbf{\texttt{test\_freq}} & \textbf{\texttt{test\_rare}} \\
        \midrule
        \multirow{4}{*}{\shortstack[l]{Dialog\\History}} 
            & \# dialogs       & \multicolumn{1}{c}{363}   & \multicolumn{1}{c}{290}   \\
            & max turn         & \multicolumn{1}{c}{16}    & \multicolumn{1}{c}{24}    \\
            & min turn         & \multicolumn{1}{c}{10}    & \multicolumn{1}{c}{10}    \\
            & avg turn         & \multicolumn{1}{c}{10.57} & \multicolumn{1}{c}{10.61} \\
        \midrule
        \multirow{4}{*}{\shortstack[l]{Question /\\Answer}}       
            & \# QA pairs     &  1,452            & 1,160              \\  
            & max dur(s)     & 13.19 / 24.80             & 19.23 / 22.11              \\
            & min dur(s)     & 2.60 / 1.11               & 2.14 / 1.30               \\
            & avg dur(s)     & 5.97 / 6.78               & 5.90 / 6.59               \\
        \bottomrule
    \end{tabular}
    \caption{Statistics of ContextDialog for Dialog History and generated QA on \texttt{test\_freq} and \texttt{test\_rare} splits. The numbers on the left and right are related to the question and answer, respectively. The term dur refers to the duration of the generated question and answer.}
    \label{tab:contextdialog_stats}
    \vskip -0.2in
\end{table}

\begin{figure*}
    \centering
    \includegraphics[width=0.95\linewidth]{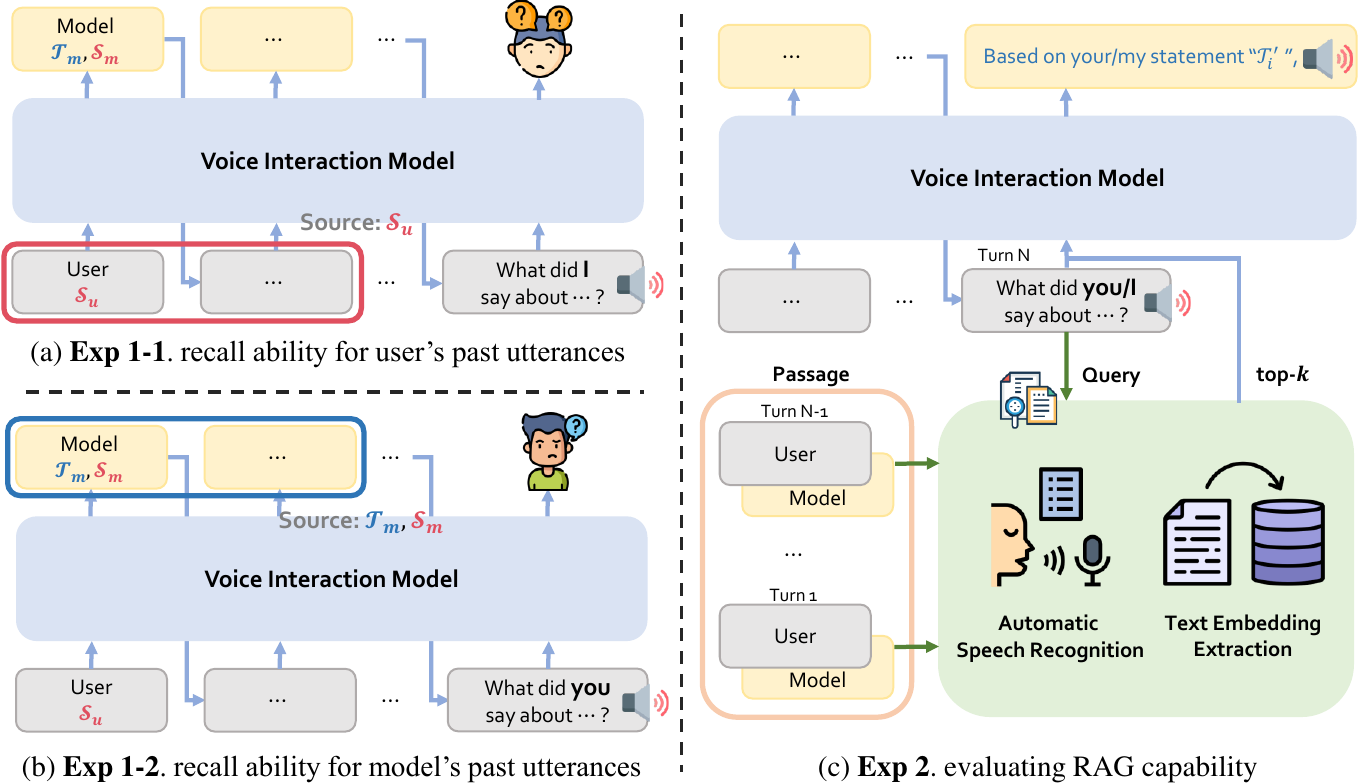}    
    \caption{Overview of our analyses. In Section \ref{subsec:recall}, we evaluate model recall by analyzing responses to questions about (a) past user and (b) past model utterances. In Section \ref{subsec:retrieval}, we examine whether (c) augmenting spoken response generation with separately retrieved utterances improves responses to questions about past utterances.}
    \label{fig_recall}
    \vskip -0.2in
\end{figure*}

\subsection{Text Question-Answer Generation}
\label{subsec:generation}
We first construct a dataset of context-recall QA pairs using \textit{gpt-4o}. Given the transcripts of MultiDialog, the model is prompted to generate questions and answers based solely on information that appeared only once in the conversation. To ensure diversity and broad applicability, we generate questions based on both user and system utterances, selecting information from either the first or second half of the conversation. This results in four QA pairs per spoken dialog. Additionally, the model is requested to output the \textit{supporting utterance}—the utterance in the conversation that serves as the clue for the answer—for each pair to enhance both data quality and usability. For a more realistic setting, the questions are designed to require detailed answers rather than simple Yes/No responses.

After generating the QA pairs, we validate each question, answer, and supporting utterance using \textit{o1-mini} \cite{openai2024openaio1card}. A validation prompt assesses their appropriateness within the dialog context through three rounds of evaluation: (1) dialog context up to just before the supporting utterance, (2) up to and including the supporting utterance, and (3) the entire conversation. The first step is to check whether the answer can be inferred without the supporting utterance, requiring a NO response, while the second and third ensure consistency across different context levels, requiring YES responses. Failed QA pairs are filtered out, and the validated pairs are used to construct the spoken QA dataset. The prompts used are in Appendix \ref{subapp:contextdialog}.

\subsection{Spoken Question-Answer Generation}
\label{subsec:tts}
To ensure that the user and the model continue naturally in the given spoken dialog, the voice of the spoken QA pair must seamlessly match that of the original conversation. To achieve this, we use Fish Speech \cite{fish-speech-v1.4}, a speaker-adaptive TTS model that generates speech in the target speaker’s timbre using a reference speech. For each QA pair, we select the longest speech segment from the original dialog for each speaker as the reference to maximize speaker similarity. To ensure accurate pronunciation, each spoken QA pair is generated five times, and the sample with the lowest word error rate (WER)—measured using a separate ASR model \cite{pmlr-v202-radford23a}—is selected. If the selected sample has a nonzero WER, it is manually reviewed, and mispronounced samples are filtered out. This process ensures that ContextDialog maintains both speaker identity and clear pronunciation in the final spoken QA pairs.

\begin{table*}[t]
\small
\centering
\renewcommand{\arraystretch}{1.2} 
\setlength{\tabcolsep}{6pt}
\begin{tabular}{l c c c c c c}
\toprule
\multirow{2}{*}{\textbf{Model}} & \multirow{2}{*}{\textbf{LLM FT}} & \multirow{2}{*}{\textbf{Modality}} & \multicolumn{3}{c}{\textbf{GPT Score}} & \multirow{2}{*}{\textbf{WER}} \\ 
\cmidrule(lr){4-6}
  & & & \textbf{User} & \textbf{System} & \textbf{Overall} & \\ 
\midrule
\rowcolor{gray!10}  &   & $\mathcal{S\rightarrow T,\colorbox{yellow!40}{\hspace{-2pt}$\underline{\bm{\mathcal{S}}}$\hspace{-2pt}}}$ & $1.94 \scalebox{0.7}{$\pm 0.07$}$   & $2.76 \scalebox{0.7}{$\pm 0.08$}$  & $2.35 \scalebox{0.7}{$\pm 0.05$}$  & 8.36\% \\ 
\rowcolor{gray!10} \multirow{-2}{*}{GLM-4-Voice \cite{glm2024chatglm}}  & \multirow{-2}{*}{\cmark} & $\mathcal{S\rightarrow \colorbox{yellow!40}{\hspace{-2pt}$\underline{\bm{\mathcal{T}}}$\hspace{-2pt}},S}$ & $2.04 \scalebox{0.7}{$\pm 0.07$}$   & $2.97 \scalebox{0.7}{$\pm 0.08$}$  & $2.50 \scalebox{0.7}{$\pm 0.06$}$  & $-$ \\ \cdashline{1-7}
glm-4-9b-chat \cite{zeng2024glm4}  &   $-$    & $\mathcal{T\rightarrow \colorbox{yellow!40}{\hspace{-2pt}$\underline{\bm{\mathcal{T}}}$\hspace{-2pt}}}$ & $4.30 \scalebox{0.7}{$\pm 0.05$}$   & $3.90 \scalebox{0.7}{$\pm 0.06$}$  & $4.10 \scalebox{0.7}{$\pm 0.04$}$  & $-$ \\ 
\midrule
\rowcolor{gray!10}         &   & $\mathcal{S\rightarrow T,\colorbox{yellow!40}{\hspace{-2pt}$\underline{\bm{\mathcal{S}}}$\hspace{-2pt}}}$ & $2.51 \scalebox{0.7}{$\pm 0.09$}$   & $3.16 \scalebox{0.7}{$\pm 0.09$}$  & $2.83 \scalebox{0.7}{$\pm 0.06$}$  & 5.90\% \\ 
\rowcolor{gray!10}   \multirow{-2}{*}{Lyra \cite{zhong2024lyra}}        & \multirow{-2}{*}{\cmark}  & $\mathcal{S\rightarrow \colorbox{yellow!40}{\hspace{-2pt}$\underline{\bm{\mathcal{T}}}$\hspace{-2pt}},S}$ & $2.67 \scalebox{0.7}{$\pm 0.09$}$   & $3.38 \scalebox{0.7}{$\pm 0.09$}$  & $3.03 \scalebox{0.7}{$\pm 0.07$}$  & $-$ \\ \cdashline{1-7}
Qwen2-VL-7B-Instruct \cite{Qwen2-VL}  &    $-$   & $\mathcal{T\rightarrow \colorbox{yellow!40}{\hspace{-2pt}$\underline{\bm{\mathcal{T}}}$\hspace{-2pt}}}$ & $3.80 \scalebox{0.7}{$\pm 0.08$}$   & $3.88 \scalebox{0.7}{$\pm 0.08$}$  & $3.84 \scalebox{0.7}{$\pm 0.06$}$  & $-$ \\ 
\midrule
\rowcolor{gray!10} &  & $\mathcal{S\rightarrow T,\colorbox{yellow!40}{\hspace{-2pt}$\underline{\bm{\mathcal{S}}}$\hspace{-2pt}}}$ & $1.73 \scalebox{0.7}{$\pm 0.06$}$   & $2.28 \scalebox{0.7}{$\pm 0.07$}$  & $2.00 \scalebox{0.7}{$\pm 0.05$}$  & 12.36\% \\ 
\rowcolor{gray!10}  \multirow{-2}{*}{Freeze-Omni \cite{xiong2024freeze}} & \multirow{-2}{*}{\xmark}   & $\mathcal{S\rightarrow \colorbox{yellow!40}{\hspace{-2pt}$\underline{\bm{\mathcal{T}}}$\hspace{-2pt}},S}$ & $2.09 \scalebox{0.7}{$\pm 0.07$}$   & $3.06 \scalebox{0.7}{$\pm 0.08$}$  & $2.57 \scalebox{0.7}{$\pm 0.06$}$  & $-$ \\ \cdashline{1-7}
Qwen2-7B-Instruct \cite{qwen2}  &   $-$    & $\mathcal{T\rightarrow \colorbox{yellow!40}{\hspace{-2pt}$\underline{\bm{\mathcal{T}}}$\hspace{-2pt}}}$ & $4.26 \scalebox{0.7}{$\pm 0.06$}$   & $3.81 \scalebox{0.7}{$\pm 0.07$}$  & $4.03 \scalebox{0.7}{$\pm 0.05$}$  & $-$ \\ 
\midrule
\rowcolor{gray!10} &  & $\mathcal{S\rightarrow T,\colorbox{yellow!40}{\hspace{-2pt}$\underline{\bm{\mathcal{S}}}$\hspace{-2pt}}}$ & $2.44 \scalebox{0.7}{$\pm 0.09$}$   & $2.84 \scalebox{0.7}{$\pm 0.09$}$  & $2.64 \scalebox{0.7}{$\pm 0.06$}$  & 24.90\% \\ 
\rowcolor{gray!10}    \multirow{-2}{*}{MiniCPM-o \cite{yao2024minicpm}}   & \multirow{-2}{*}{\cmark}   & $\mathcal{S\rightarrow \colorbox{yellow!40}{\hspace{-2pt}$\underline{\bm{\mathcal{T}}}$\hspace{-2pt}},S}$ & $3.22 \scalebox{0.7}{$\pm 0.09$}$   & $3.93 \scalebox{0.7}{$\pm 0.08$}$  & $3.58 \scalebox{0.7}{$\pm 0.06$}$  & $-$ \\ \cdashline{1-7}
Qwen2.5-7B-Instruct \cite{qwen2.5}  &  $-$     & $\mathcal{T\rightarrow \colorbox{yellow!40}{\hspace{-2pt}$\underline{\bm{\mathcal{T}}}$\hspace{-2pt}}}$ & $4.28 \scalebox{0.7}{$\pm 0.05$}$   & $3.84 \scalebox{0.7}{$\pm 0.06$}$  & $4.06 \scalebox{0.7}{$\pm 0.04$}$  & $-$ \\ 
\bottomrule
\end{tabular}
\vskip -0.05in
\caption{Evaluation results for voice interaction models, including the instruct fine-tuned version of each model's backbone LLM. $\mathcal{S}$ and $\mathcal{T}$ represent speech and text, respectively. The bold and underlined modality indicates the data type used for evaluation. ``Modality'' indicates input → output data type. ``LLM FT'' shows whether the backbone LLM was fine-tuned during training. ``User'' and ``System'' represent scores for responses to past user and model utterances, respectively. ``Overall'' denotes the score across all responses. ``WER'' refers to the word error rate between the model’s intermediate text response and the transcribed spoken response, highlighting degradation from speech synthesis. GPT Scores are reported with a 95\% confidence interval.}
\label{tab:recall}
\vskip -0.2in
\end{table*}

\section{Experiments}
\label{sec:experiments}

\begin{figure*}
    \centering
    \includegraphics[width=0.95\linewidth]{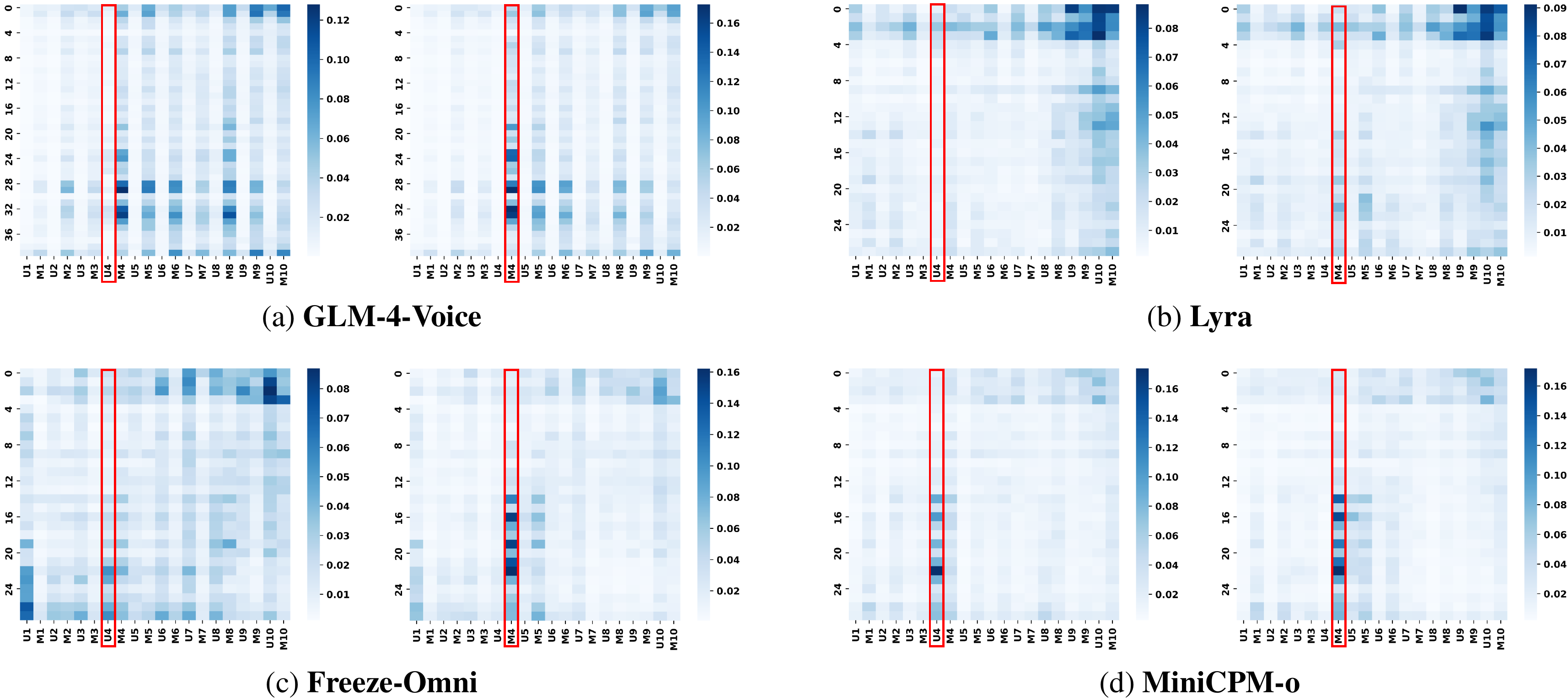}    
    \vskip -0.05in
    \caption{Attention maps for ground truth answers given each model's past dialog and question. The $x$-axis of the figure indicates the order of utterances of each speaker (``U'' for user, ``M'' for model), while the $y$-axis shows the index of attention layer. In each subfigure, the left side represents questions about past user utterances, and the right side represents questions about past model utterances. Red boxes indicate the positions of supporting utterances.}
    \label{fig_attn_map}
    \vskip -0.2in
\end{figure*}

In this section, we present the results of two experiments and analyses using ContextDialog. In Section \ref{subsec:recall}, we demonstrate that open-source voice interaction models struggle to recall past information on their own, particularly user-specific information that exists solely in spoken form. Then, in Section \ref{subsec:retrieval}, we show that even when leveraging a more advanced dedicated text retriever, models fail to respond robustly given the retrieved information, yielding limited improvements in spoken response generation. These two analyses highlight a critical yet often overlooked aspect of voice interaction models, their ability to remember past interactions, which is essential for real-world deployment.

For our experiments, we select four open-source multi-turn voice interaction models: GLM-4-Voice \cite{zeng2024glm4}, MiniCPM-o 2.6 \cite{yao2024minicpm}, Freeze-Omni \cite{xiong2024freeze}, and Lyra \cite{zhong2024lyra}. To support real-time generation and minimize latency, these models generate responses directly from the input speech without an intermediate speech-to-text conversion. To mitigate semantic degradation in speech-only generation, these models generate text responses alongside spoken responses: GLM-4-Voice employs an interleaved token generation approach, alternating between text and speech tokens (Figure \ref{fig_decoder}(a)), while MiniCPM-o, Freeze-Omni, and Lyra generate text responses while simultaneously synthesizing speech using real-time generated text tokens and the LLM’s hidden states (Figure \ref{fig_decoder}(b)). 

In all experiments, we evaluate each model's spoken response using the LLM-as-a-judge approach~\cite{NEURIPS2023_91f18a12}, following previous works~\cite{chen2024slamomnitimbrecontrollablevoiceinteraction, zeng2025scaling}. This setting is denoted as $\mathcal{S} \rightarrow \mathcal{T},\, \colorbox{yellow!40}{\hspace{-2pt}$\underline{\bm{\mathcal{S}}}$\hspace{-2pt}}$, where $\mathcal{S}$ refers to speech data and $\mathcal{T}$ to text data. The bold and underlined modality symbol indicates the evaluation target in each configuration. We employ \textit{gpt-4o-mini} for evaluation, referred to as the GPT Score in this paper, using a five-point scale where higher scores indicate better performance. We design prompts to assess recall by measuring how well the generated responses contain the ground truth information relevant to the given question, as detailed in Appendix~\ref{subapp:evaluation}. Since \textit{gpt-4o-mini} is tailored to text inputs, we first convert the spoken responses into text using \textit{whisper-large-v3}~\cite{pmlr-v202-radford23a}.

Additionally, considering that each model also generates an intermediate text response corresponding to the spoken output, we also evaluate it ($\mathcal{S\rightarrow}\colorbox{yellow!40}{\hspace{-2pt}$\underline{\bm{\mathcal{T}}}$\hspace{-2pt}}\mathcal{,S}$). By analyzing the evaluation results along with the word error rate (WER) between the text response and the transcribed spoken response, we can disentangle recall ability from speech synthesis capability, allowing us to identify cases where the model successfully recalls information but fails in speech synthesis.

We use the official implementations, hyperparameters, and checkpoints for all four models (Section \ref{subapp:licenses}), running experiments on a single NVIDIA A40 GPU. Detailed model descriptions and additional analyses are provided in Appendix \ref{subapp:models}.

\subsection{Does Your Model Truly Recall Past Information?}
\label{subsec:recall}

Using ContextDialog, we examine whether these models can recall or remind users of past utterances, either from the user or the model itself. To assess differences in question difficulty, we additionally evaluate the instruct fine-tuned versions of each model’s backbone LLM \cite{glm2024chatglm, qwen2, qwen2.5, Qwen2-VL}, providing a basis for comparing the difficulty of questions based on user-spoken versus model-generated utterances. 

The scores for each model on questions about past user utterances and the model’s own responses, along with their 95\% confidence intervals and averages, are presented in Table \ref{tab:recall}. In this table, we observe two key patterns. First, in multi-turn dialogs requiring past context, all voice interaction models (shaded in gray) show significant performance drop compared to text-based counterparts (unshaded), regardless of whether evaluation is on the intermediate text response ($\mathcal{S\rightarrow}\colorbox{yellow!40}{\hspace{-2pt}$\underline{\bm{\mathcal{T}}}$\hspace{-2pt}}\mathcal{,S}$) or the transcribed response ($\mathcal{S\rightarrow T,\colorbox{yellow!40}{\hspace{-2pt}$\underline{\bm{\mathcal{S}}}$\hspace{-2pt}}}$). This degradation is particularly pronounced in Freeze-Omni, where the LLM is frozen during speech model training (LLM FT: \xmark). These results indicate that expanding a pre-trained LLM to speech significantly weakens its ability to process long contexts.

Secondly, unlike their text-based counterparts (unshaded), voice interaction models (shaded in gray) perform consistently better when recalling their own past utterances than the user’s ($p<0.01$). This stems from the generation mechanism of recent voice interaction models. Since speech-only output degrades semantic modeling, most modern models generate text alongside speech to leverage the backbone LLM’s text capability. Consequently, when responding to questions about the model’s past utterances, both text and speech are utilized (Figure \ref{fig_recall}(b)), whereas for user utterances, the model must rely solely on speech (Figure \ref{fig_recall}(a)).

\begin{figure*}
    \centering
    \includegraphics[width=0.98\linewidth]{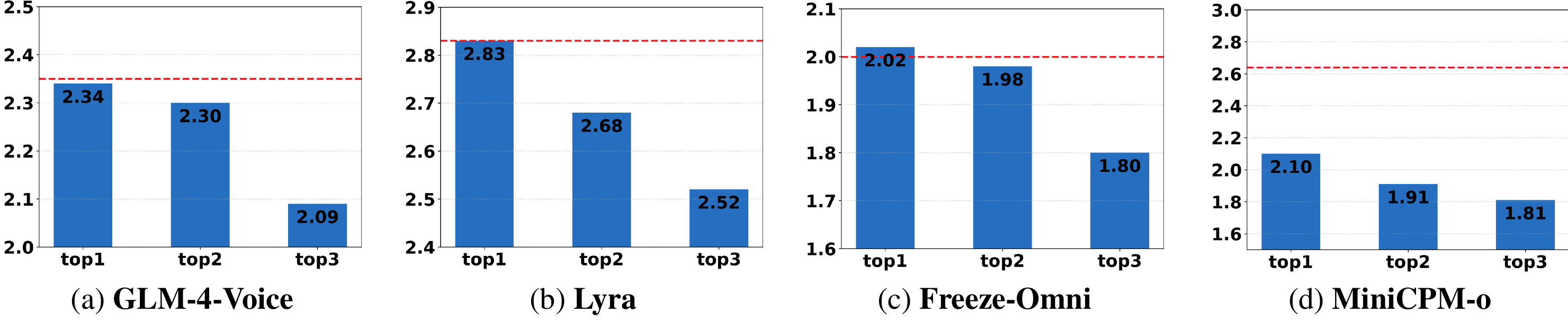}    
    \vskip -0.05in
    \caption{The results of applying the RAG method to each model are shown. The $y$-axis values indicate GPT Scores on a 5-point scale, with higher scores representing better performance. The red dashed line indicates the results generated without RAG (Section \ref{subsec:recall}). The evaluation is based on the transcribed spoken response $\mathcal{S\rightarrow T,\colorbox{yellow!40}{\hspace{-2pt}$\underline{\bm{\mathcal{S}}}$\hspace{-2pt}}}$.}
    \label{fig_rag_results}
    \vskip -0.2in
\end{figure*} 

We further analyze how models respond to questions about past user and model utterances by examining their attention maps during response generation as shown in Figure \ref{fig_attn_map}. The horizontal axis represents the turn index (``U'' for user, ``M'' for model), and the vertical axis represents the attention layer index. We sum attention weights over all tokens in each utterance. As shown, models attend less to supporting utterances when answering questions about past user utterances than model utterances. This suggests that an inherent bias, where models allocate less attention to user-spoken content, contributes to the recall gap and highlights the need for improved modeling capabilities.

The findings in this section reveal modality-specific differences, both compared to text interaction models and within speech models. They underscore the need to improve voice interaction models by introducing training and generation methods to better utilize long-range conversational history. In the following section, we validate a practical approach to enhancing past information utilization with minimal computational cost when the model fails to recall relevant details. Specifically, we examine the effectiveness of retrieval-augmented generation (RAG) in voice interaction models.

\subsection{Does Your Model Reliably Augment Retrieved Information into Generation?}
\label{subsec:retrieval}

Leveraging RAG methods from the NLP domain \cite{JMLR:v24:23-0037, NEURIPS2020_6b493230}, we assess whether voice interaction models can effectively utilize past utterances when retrieved by a dedicated module, as illustrated in Figure \ref{fig_recall}(c). Given our observation in Section \ref{subsec:recall} that models struggle more with speech than text, we transcribe past user and model utterances at the end of each speech segment using a separate ASR model. These transcriptions are stored with their corresponding text embeddings, extracted via a separate retriever.

Once stored, these transcriptions serve as passages from which relevant information can be retrieved when a user query arrives. Upon receiving input speech, we convert it into text using the same ASR model, extract its embedding, and retrieve the top-\textit{k} most relevant past utterances by comparing cosine similarity. These retrieved utterances are then augmented into spoken response generation. We use \textit{whisper-large-v3-turbo} \cite{pmlr-v202-radford23a} for the ASR model and \textit{e5-large-v2} \cite{Wang2022TextEB}, a widely used retrieval model in NLP.

The retrieved texts are incorporated into the generation stage using the following format:  
\texttt{Based on your/my statement: ``RETRIEVED TRANSCRIBED TEXT1'', your/my statement: ``RETRIEVED TRANSCRIBED TEXT2'' ...}. The choice between \texttt{your} and \texttt{my} depends on the speaker of the retrieved utterance, ensuring clear integration into the prompt. The model then utilizes this prompt to generate spoken responses, as shown in Figure \ref{fig_recall}(c). Details on how each model incorporates this prompt into spoken response generation are in Appendix \ref{subsubapp:models_for_analysis}, while experiments with various other prompts are discussed in Appendix \ref{subsubapp:retrieval_prompt}.

\begin{table}[t]
\small
\centering
\begin{tabular}{l|c|c}
\toprule
\textbf{Model}               & \textbf{Prompt} & \textbf{GPT Score}                                \\ \midrule
\multirow{3}{*}{GLM-4-Voice} &  \xmark             & $2.35\pm 0.05$ \\ \cmidrule{2-3} 
                             & Supporting      & $\mathbf{2.60\pm 0.05}$ \\ \cmidrule{2-3} 
                             & Irrelevant      & $1.87\pm 0.05$ \\ \midrule
\multirow{3}{*}{Lyra}        & \xmark               & $2.83\pm 0.06$ \\ \cmidrule{2-3} 
                             & Supporting      & $\mathbf{3.44\pm 0.05}$ \\ \cmidrule{2-3} 
                             & Irrelevant      & $1.96\pm 0.05$ \\ \midrule
\multirow{3}{*}{Freeze-Omni} &\xmark              & $2.00\pm 0.05$ \\ \cmidrule{2-3} 
                             & Supporting      & $\mathbf{2.38\pm 0.04}$ \\ \cmidrule{2-3} 
                             & Irrelevant      & $1.54\pm 0.04$ \\ \midrule
\multirow{3}{*}{MiniCPM-o}   & \xmark               & $\mathbf{2.64\pm 0.06}$ \\ \cmidrule{2-3} 
                             & Supporting      & $2.49\pm 0.06$ \\ \cmidrule{2-3} 
                             & Irrelevant      & $1.63\pm 0.05$ \\
                             \bottomrule
\end{tabular}
\vskip -0.05in
\caption{GPT Score results when augmenting spoken response generation with either the ground-truth supporting utterance (``Supporting'') or an entirely unrelated utterance (``Irrelevant'') as prompts.}
\label{tab:rag_support}
\vskip -0.2in
\end{table}

\begin{table*}[t]
\small
\centering
\renewcommand{\arraystretch}{1.2} 
\setlength{\tabcolsep}{6pt}
\begin{tabular}{l c c c c c}
\toprule
\multirow{2}{*}{\textbf{Model}} & \multirow{2}{*}{\textbf{Retriever}} & \multirow{2}{*}{\textbf{ASR}} & \multicolumn{3}{c}{\textbf{GPT Score}} \\ 
\cmidrule(lr){4-6}
  & & & \textbf{top-\textit{1}} & \textbf{top-\textit{2}} & \textbf{top-\textit{3}} \\ 
\midrule
\rowcolor{gray!10} \cellcolor{white} &   & \cmark & $2.34 \scalebox{0.7}{$\pm 0.05$}$   & $2.30 \scalebox{0.7}{$\pm 0.05$}$  & $2.09 \scalebox{0.7}{$\pm 0.05$}$  \\ 
\rowcolor{gray!10} \cellcolor{white}GLM-4-Voice \cite{glm2024chatglm}  & \multirow{-2}{*}{\textit{e5-large-v2} \cite{Wang2022TextEB}} & \xmark & $2.42 \scalebox{0.7}{$\pm 0.05$}$   & $2.40 \scalebox{0.7}{$\pm 0.05$}$  & $2.15 \scalebox{0.7}{$\pm 0.05$}$  \\ \cdashline{2-6}
  &   SONAR \cite{Duquenne:2023:sonar_arxiv}    & $-$ & $2.24 \scalebox{0.7}{$\pm 0.05$}$   & $2.15 \scalebox{0.7}{$\pm 0.05$}$  & $1.97 \scalebox{0.7}{$\pm 0.05$}$  \\ 
\midrule
\rowcolor{gray!10} \cellcolor{white}&   & \cmark & $2.83 \scalebox{0.7}{$\pm 0.06$}$   & $2.68 \scalebox{0.7}{$\pm 0.06$}$  & $2.52 \scalebox{0.7}{$\pm 0.06$}$  \\ 
\rowcolor{gray!10}  \cellcolor{white} Lyra \cite{zhong2024lyra}       & \multirow{-2}{*}{\textit{e5-large-v2} \cite{Wang2022TextEB}}  & \xmark & $2.94 \scalebox{0.7}{$\pm 0.06$}$   & $2.78 \scalebox{0.7}{$\pm 0.06$}$  & $2.68 \scalebox{0.7}{$\pm 0.06$}$  \\ \cdashline{2-6}
  &    SONAR \cite{Duquenne:2023:sonar_arxiv}   & $-$ & $2.48 \scalebox{0.7}{$\pm 0.06$}$   & $2.39 \scalebox{0.7}{$\pm 0.06$}$  & $2.25 \scalebox{0.7}{$\pm 0.06$}$  \\ 
\midrule
\rowcolor{gray!10} \cellcolor{white} &  & \cmark & $2.02 \scalebox{0.7}{$\pm 0.04$}$   & $1.98 \scalebox{0.7}{$\pm 0.04$}$  & $1.80 \scalebox{0.7}{$\pm 0.04$}$  \\ 
\rowcolor{gray!10} \cellcolor{white} Freeze-Omni \cite{xiong2024freeze} & \multirow{-2}{*}{\textit{e5-large-v2} \cite{Wang2022TextEB}}   & \xmark & $2.08 \scalebox{0.7}{$\pm 0.04$}$   & $2.03 \scalebox{0.7}{$\pm 0.04$}$  & $1.90 \scalebox{0.7}{$\pm 0.04$}$  \\ \cdashline{2-6}
  &   SONAR \cite{Duquenne:2023:sonar_arxiv}    & $-$ & $1.83 \scalebox{0.7}{$\pm 0.04$}$   & $1.77 \scalebox{0.7}{$\pm 0.04$}$  & $1.67 \scalebox{0.7}{$\pm 0.04$}$  \\ 
\midrule
\rowcolor{gray!10} \cellcolor{white} &  & \cmark & $2.10 \scalebox{0.7}{$\pm 0.05$}$   & $1.91 \scalebox{0.7}{$\pm 0.05$}$  & $1.81 \scalebox{0.7}{$\pm 0.05$}$  \\ 
\rowcolor{gray!10} \cellcolor{white}   MiniCPM-o \cite{yao2024minicpm}   & \multirow{-2}{*}{\textit{e5-large-v2} \cite{Wang2022TextEB}}   & \xmark & $2.16 \scalebox{0.7}{$\pm 0.06$}$   & $1.98 \scalebox{0.7}{$\pm 0.05$}$  & $1.86 \scalebox{0.7}{$\pm 0.05$}$  \\ \cdashline{2-6}
  &  SONAR \cite{Duquenne:2023:sonar_arxiv}     & $-$ & $2.01 \scalebox{0.7}{$\pm 0.05$}$   & $1.82 \scalebox{0.7}{$\pm 0.05$}$  & $1.78 \scalebox{0.7}{$\pm 0.05$}$  \\ 
\bottomrule
\end{tabular}
\vskip -0.05in
\caption{Evaluation results for RAG with voice interaction models. `ASR'' indicates whether RAG is performed using ASR-transcribed text (\cmark) or ground-truth text (\xmark). The scores are reported with a 95\% confidence interval.}
\label{tab:rag_wo_asr}
\vskip -0.1in
\end{table*}

The experimental results on integrating RAG into voice interaction models are presented in Figure \ref{fig_rag_results}, where (a)–(d) correspond to the four evaluated models. The red dashed line indicates baseline performance when models generate responses based solely on intrinsic recall without RAG (Section \ref{subsec:recall}). These results are measured using the ASR transcript of the spoken response ($\mathcal{S\rightarrow T,\colorbox{yellow!40}{\hspace{-2pt}$\underline{\bm{\mathcal{S}}}$\hspace{-2pt}}}$) for all QA pairs, and trends in the intermediate text response are similar, as detailed in Appendix \ref{subsubapp:results_on_intermediate_text_responses}.

As shown in the figure, all models perform similarly or worse with RAG, showing little to no improvement as the number of retrieved utterances increases. We attribute this to two main factors. First, while RAG increases the chances of retrieving and using supporting utterances, retrieval failures introduce irrelevant sentences that add noise and disrupt generation. Second, unlike text-based models, voice interaction models are generally trained to avoid long responses, as users do not expect lengthy monologs. However, RAG adds prompts to the generation process, leading to longer responses that contradict the models' training tendencies.

\subsubsection{Analyses}
We observe that incorporating utterances retrieved by a dedicated retrieval module \cite{Wang2022TextEB} into spoken response generation has little effect on voice interaction models. To further investigate this phenomenon, we conduct various experiments.

To determine whether prompting itself is ineffective for voice interaction models, we conduct two experiments: (1) providing the supporting utterance from the ContextDialog QA pair as a prompt instead of retrieved utterances and (2) using an unrelated utterance as a prompt to generate the spoken response. As in previous evaluations, we assess the spoken response based on its transcribed text ($\mathcal{S\rightarrow T,\colorbox{yellow!40}{\hspace{-2pt}$\underline{\bm{\mathcal{S}}}$\hspace{-2pt}}}$), with results shown in Table \ref{tab:rag_support}.

For models other than MiniCPM-o, we observe that providing the correct supporting utterance improves performance on QA requiring past information, while using an incorrect utterance as a prompt degrades performance. This suggests that for most models, the primary obstacles to using RAG for remembering past conversations in voice interaction models lie not in the act of augmentation itself, but in factors beyond incorporating relevant information, such as retrieval errors.

To examine whether the limited effectiveness of RAG is primarily due to ASR errors, we analyze the impact of recognition errors in retrieving past utterances. Specifically, we compare two approaches: (1) retrieving using the ground-truth text of past conversations and the ground-truth transcript of the input speech and (2) retrieving directly from speech with a speech retriever module, bypassing the recognition process.

Since no suitable open-source speech retriever module is available, we use SONAR \cite{Duquenne:2023:sonar_arxiv}, which, while not primarily designed for retrieval, extracts semantic embeddings from speech and retrieves past utterances based on cosine similarity. Note that since the voice interaction models rely on text for spoken response generation, retrieved information is provided in text form regardless of retriever modality.

\begin{table}[]
\small
\centering
\begin{tabular}{l|c|ccc}
\toprule
\multirow{2}{*}{\textbf{Retriever}}                    & \multirow{2}{*}{\textbf{ASR}} & \multicolumn{3}{c}{\textbf{Recall}}                                                                                                                                      \\ \cmidrule{3-5} 
                                                       &                               & \multicolumn{1}{c|}{\textbf{top-\textit{1}}} & \multicolumn{1}{c|}{\textbf{top-\textit{2}}} & \textbf{top-\textit{3}} \\ \midrule
\multirow{2}{*}{\textit{e5-large-v2}} & \cmark         & \multicolumn{1}{c|}{0.5773}                                   & \multicolumn{1}{c|}{0.7339}                                   & 0.7959                                   \\ \cmidrule{2-5} 
                                                       & \xmark         & \multicolumn{1}{c|}{0.5827}                                   & \multicolumn{1}{c|}{0.7561}                                   & 0.8212                                   \\ \midrule
SONAR                                                  & $-$                           & \multicolumn{1}{c|}{0.3955}                                   & \multicolumn{1}{c|}{0.5306}                                   & 0.6087    \\
\bottomrule
\end{tabular}
\vskip -0.05in
\caption{Retrieval performance for each model used in the analysis, measuring the probability of the supporting utterance being included in the top-\textit{k} utterances. ``ASR'' indicates that retrieval is performed using transcripts obtained from the ASR model.}
\label{tab:retriever_recall}
\vskip -0.2in
\end{table}

As shown in Table \ref{tab:rag_wo_asr}, ASR has minimal impact on RAG performance for text-based retrievers (``ASR'' \cmark vs. \xmark). In contrast, using a speech retriever leads to a relatively significant performance drop. These results align with the retrieval performance in Table \ref{tab:retriever_recall}, where ASR does not substantially affect the retriever’s ability to include the supporting utterance in the top-\textit{k} results. Additionally, the speech retriever is not originally designed for retrieval, and training challenges—such as longer audio sequences and limited data—contribute to recall degradation, leading to performance decline.

The observations in Section \ref{subsec:recall} highlight the recall difficulty of speech being substantial compared to text. The findings in this section show that even when models retrieve information through an external module and augment it into generation, they fail to use it effectively, suggesting two key areas for improvement. First, even when explicitly provided with the supporting utterance, current models underperform compared to text-based counterparts, underscoring the need for stronger conversational capabilities in voice interaction models. Second, while several methods were developed to ensure robustness against retrieval noise in the NLP domain \cite{Chen_Lin_Han_Sun_2024, yoran2024making}, voice interaction models require better training and inference strategies to enhance resilience to retrieval noise alongside general modeling improvements.

\section{Conclusion}
In this work, we conducted an in-depth analysis of a critical yet underexplored challenge in open-source voice interaction models: maintaining and utilizing past utterances. To address the lack of benchmarks that explicitly require accurate reference to past dialog, we introduced ContextDialog, a speech-to-speech benchmark designed to systematically evaluate a model’s ability to recall utterances from previous turns. Using this benchmark, our experiments revealed that models struggle with recalling past utterances and remain highly sensitive to retrieval errors, limiting improvements even with dedicated retriever. These findings highlight a crucial gap in memory retention for open-source models, emphasizing the need for stronger conversational memory methods, such as improved long-context modeling, robust RAG techniques, or dedicated memory modules. We hope that our work may act as a trigger to raise awareness to this overlooked challenge and encourage future research to further enhance the usability and effectiveness of voice interaction models.
\section*{Acknowledgments}
This work was supported by the National Research Foundation of Korea (NRF) grant funded by the Korea government (MSIT) [No. 2022R1A3B1077720], Institute of Information \& Communications Technology Planning \& Evaluation (IITP) grant funded by the Korea government (MSIT) [NO. RS-2021-II211343, Artificial Intelligence Graduate School Program (Seoul National University), NO. RS-2022-II220959], the BK21 FOUR program of the Education and Research Program for Future ICT Pioneers, Seoul National University in 2024, Samsung Electronics (IO221213-04119-01), and a grant from the Yang Young Foundation.

\section*{Limitations}
Our study highlights the overlooked issue of history recall in voice interaction models and introduces a benchmark for systematic evaluation. We focus on open-source multi-round voice interaction models, analyzing them with additional results in Appendix \ref{subsubapp:additional_models}. However, other open-source models not covered in our analyses may exist. Additionally, while most recent models enhance semantic modeling by jointly generating spoken and text responses, some still generate speech directly without relying on intermediate text. Future research could extend our analysis to these models.

Another limitation of our retrieval-based analyses is its focus on text-based retrieval-augmented generation (RAG). Currently, no well-established speech retriever modules exist for open-source models, and open-source voice interaction models struggle with speech-based prompting for RAG, restricting our analysis to text prompts. Furthermore, we do not consider the latency introduced by RAG. Developing low-latency speech retrievers that efficiently integrate spoken information—including linguistic and non-verbal cues—remains crucial for real-time conversational applications.

In addition, our study has a limitation that the proposed benchmark, ContextDialog, is synthetic. While both LLM-based and human verification are applied, the question-answer pair is generated using \textit{gpt-4o}, and the corresponding audio is synthesized using a separate text-to-speech (TTS) model. Despite this, many recent benchmarks for evaluation and datasets for training are similarly constructed using LLM APIs, and the growing quality of TTS models has made synthetic data increasingly common in the speech community. Nonetheless, real-world data remains more valuable in terms of robustness and alignment with practical use cases. As a next step, we plan to extend our work to real-world, human-collected datasets involving multi-turn voice assistant scenarios, where models must effectively leverage prior context and the audio is recorded by real-world users.

Finally, our benchmark addresses only the simplest form of questions requiring past information, those directly retrieving and utilizing prior context. While our analysis shows that current open-source voice interaction models struggle even with basic recall, more advanced benchmarks will be necessary as these models evolve. For instance, future benchmarks could move beyond simple retrieval-based responses to questions requiring deeper reasoning over past context. Additionally, a benchmark focusing on memory capabilities in common voice interaction scenarios—such as handling fragmented information (e.g., a customer providing a phone number in segments)—would be valuable for assessing more complex recall abilities.

\section*{Ethical Considerations}

Our analysis highlights the recall capabilities of voice interaction models, particularly in personalized voice assistants that rely on past interactions for customized services. However, this capability inherently raises security and privacy concerns, as stored conversational data may be vulnerable to unauthorized access. As voice assistants become more deeply integrated into daily life, ensuring they retain necessary context while safeguarding user data is crucial. Therefore, alongside advancements in memory retention and utilization, developing robust mechanisms to protect stored history must remain a parallel research priority.

\bibliography{acl_latex}

\appendix
\clearpage
\section{Appendix}
\label{app:appendix}

\subsection{Additional Details and Analysis}
\label{subapp:models}

We provide a more detailed description of the statistics of ContextDialog in Section~\ref{subsubapp:datasets_for_analysis}. Section~\ref{subsubapp:models_for_analysis} introduces the four models used in our experiments—GLM-4-Voice~\cite{zeng2024glm4}, Freeze-Omni~\cite{xiong2024freeze}, Lyra~\cite{zhong2024lyra}, and MiniCPM-o~\cite{yao2024minicpm}—along with how the text-based RAG method described in Section~\ref{subsec:retrieval} is applied to each. Section~\ref{subsubapp:results_on_intermediate_text_responses} presents results on intermediate text responses for questions about past utterances, which were not covered in Section~\ref{subsec:retrieval}. Sections \ref{subsubapp:results_on_closed_source models} and \ref{subsubapp:human evaluation results} provide additional GPT Score comparisons between closed-source and open-source voice interaction models, as well as human evaluation results that extend the main experiments. Section~\ref{subsubapp:additional_models} discusses other open-source models and explains the rationale for excluding certain ones. We also include additional experiments on datasets and retrieval prompts in Sections~\ref{subsubapp:additional_datasets} and~\ref{subsubapp:retrieval_prompt}. Finally, Section~\ref{subsubapp:failure_cases} categorizes and illustrates failure cases in which models struggle to handle questions related to past utterances.

\subsubsection{Dataset Details}
\label{subsubapp:datasets_for_analysis}
For clarity and detailed understanding, we present the dataset statistics of our proposed benchmark, ContextDialog, in Table~\ref{tab:contextdialog_app}, broken down by data split (``Split''), the source of the recalled information (``Spk'') and the type of utterance (``Type'').

\begin{table}[t]
\small
\centering
\renewcommand{\arraystretch}{1.2}
\setlength{\tabcolsep}{6pt}
\begin{tabular}{l c c c c c}
\toprule
\textbf{Split} & \textbf{Spk} & \textbf{Type} & \textbf{Max} & \textbf{Min} & \textbf{Avg} \\
\midrule
\multirow{4}{*}{\texttt{test\_freq}} 
    & \multirow{2}{*}{$\mathcal{S}$} & $\mathcal{Q}$ & $13.19$ & $2.60$ & $6.03$ \\
    &                                & $\mathcal{A}$ & $24.80$ & $1.11$ & $6.48$ \\
\cdashline{2-6}
    & \multirow{2}{*}{$\mathcal{U}$} & $\mathcal{Q}$ & $12.91$ & $2.97$ & $5.92$ \\
    &                                & $\mathcal{A}$ & $18.85$ & $1.58$ & $7.08$ \\
\midrule
\multirow{4}{*}{\texttt{test\_rare}} 
    & \multirow{2}{*}{$\mathcal{S}$} & $\mathcal{Q}$ & $19.23$ & $2.14$ & $6.01$ \\
    &                                & $\mathcal{A}$ & $20.94$ & $1.30$ & $6.46$ \\
\cdashline{2-6}
    & \multirow{2}{*}{$\mathcal{U}$} & $\mathcal{Q}$ & $11.15$ & $2.83$ & $5.79$ \\
    &                                & $\mathcal{A}$ & $22.11$ & $1.39$ & $6.72$ \\
\bottomrule
\end{tabular}
\vskip -0.05in
\caption{Statistics of ContextDialog for generated QA on the \texttt{test\_freq} and \texttt{test\_rare} splits. The column ``Spk'' indicates the speaker of the recalled utterance ($\mathcal{S}$: system, $\mathcal{U}$: user), and ``Type'' denotes the role of each utterance in the QA pair, with $\mathcal{Q}$ representing the question and $\mathcal{A}$ the answer. ``Max'', ``Min'', and ``Avg'' refer to the maximum, minimum, and average utterance lengths, respectively, measured in seconds.}
\label{tab:contextdialog_app}
\vskip -0.2in
\end{table}

\begin{figure}
    \centering
    \includegraphics[width=0.95\linewidth]{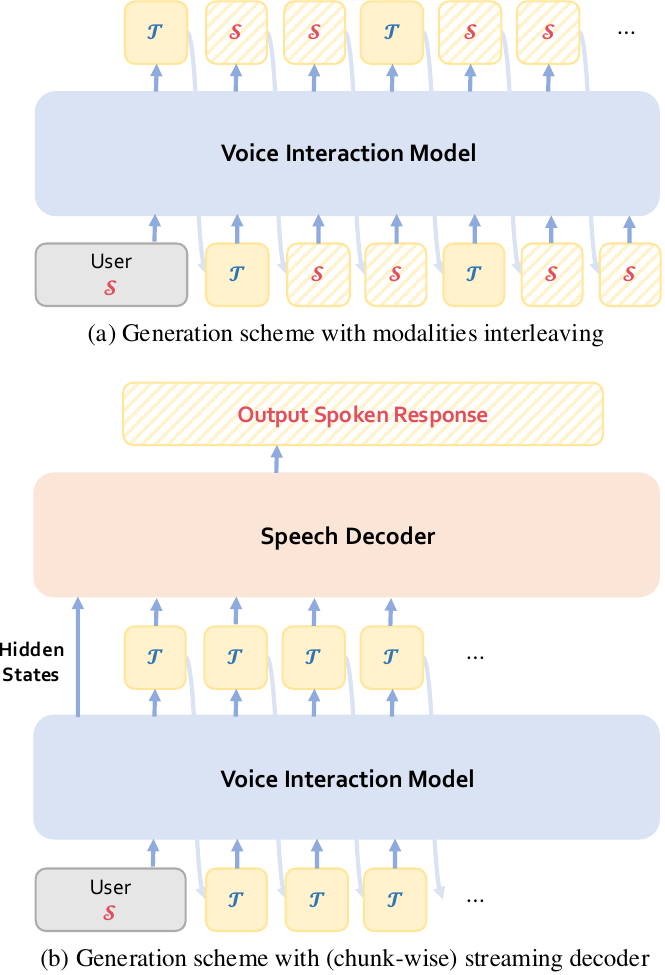}    
    \caption{Two representative approaches for generating text alongside spoken responses to enhance semantic coherence in voice interaction models.}
    \label{fig_decoder}
    \vskip -0.2in
\end{figure}

\subsubsection{Model Details}
\label{subsubapp:models_for_analysis}

\textbf{(1) GLM-4-Voice} tokenizes raw waveforms into discrete tokens, enabling training with a pre-trained LLM to construct a cross-modal spoken dialog model using both speech and text tokens. The speech tokenization module incorporates a pooling layer and a vector quantization layer \cite{NIPS2017_7a98af17} into the pre-trained \textit{whisper} encoder \cite{pmlr-v202-radford23a}, modifying it to be causal with block-wise causal attention for streaming support. For token-to-speech reconstruction, the model employs a CosyVoice-based module \cite{du2024cosyvoicescalablemultilingualzeroshot} with chunk-wise autoregressive modeling. It is trained to generate speech tokens in response to input speech tokens while also generating text tokens to leverage the LLM’s text capability. To minimize latency, instead of generating the full text sequence before speech, it adopts interleaved generation, alternating 13 text tokens with 26 speech tokens per step (Figure \ref{fig_decoder}(a)).

\begin{figure*}[t]
    \centering
    \includegraphics[width=0.98\linewidth]{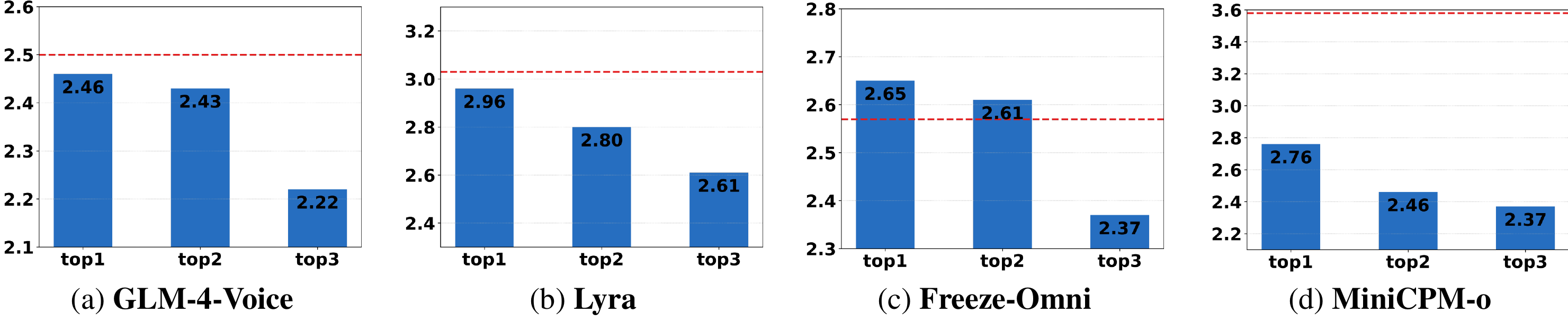}    
    \caption{The results of applying a RAG method to each model are shown. The red dashed line indicates the results generated without RAG (Section \ref{subsec:recall}). The evaluation is based on the intermediate text response $\mathcal{S\rightarrow}\colorbox{yellow!40}{\hspace{-2pt}$\underline{\bm{\mathcal{T}}}$\hspace{-2pt}}\mathcal{,S}$.}
    \label{fig_rag_results_app}
    \vskip -0.2in
\end{figure*}

In Section \ref{subsec:retrieval}, we use the pre-trained retriever \textit{e5-large-v2} \cite{Wang2022TextEB} to select the \textit{top-k} utterances as prompts to evaluate RAG in spoken response generation. Retrieved sentences are formatted using the prompt template \texttt{Based on your/my statement ``...''} and tokenized. Since the prompt typically exceeds 13 tokens, it is sequentially fed into the model at each text generation step, filling the text token slots in GLM-4-Voice's interleaved generation process (13 text tokens alternating with 26 speech tokens) until fully consumed. The model generates response text only after completing the prompt, and for speech token slots, it first produces speech tokens corresponding to the prompt before generating tokens for the newly generated text. As both the intermediate text response and the transcribed spoken response contain the prompt, we remove it using \textit{gpt-4o} before final evaluation to ensure a fair comparison.

\textbf{(2) Freeze-Omni} is built by freezing the backbone LLM and training only the plug-in speech encoder and decoder, without additional speech tokenization. Input speech is processed through a separately trained ASR encoder, which supports chunk-wise streaming by feeding encoder outputs in segments. These outputs pass through an adapter before entering the frozen LLM, where speech features are converted into LLM-compatible inputs to generate text responses. The plug-in speech decoder then takes the text response and the LLM’s hidden states to generate speech alongside text. This design preserves the LLM’s text capabilities while enabling speech generation through dedicated encoding and decoding modules.

Freeze-Omni, along with Lyra and MiniCPM-o, generates speech using text output, hidden states, or both, as illustrated in Figure \ref{fig_decoder}(b), with speech decoding typically performed end-to-end. To integrate RAG, past transcribed utterances retrieved by a separate model are formatted according to a predefined template and provided as a prefix before text response generation. That is, in addition to past conversational history and the user's input speech, the model utilizes the retrieved prompt as a prefix to generate the intermediate text response. However, the prefix and corresponding hidden states are excluded from the decoder input; only the subsequently generated response is used, ensuring that the spoken response does not include the prompt.

\textbf{(3) Lyra} is an Omni model capable of processing text, speech, and visual data such as video and images. For speech, it employs \textit{whisper-large-v3} as an encoder to extract information for the backbone LLM, while its speech decoder is trained similarly to LLaMA-Omni \cite{fang2025llamaomni}. Lyra generates spoken responses using discrete units obtained via \textit{k}-means clustering on intermediate representations from a self-supervised model \cite{10.1109/TASLP.2021.3122291}. Given user speech features as input, Lyra generates text alongside discrete speech units. While generating text responses, the LLM’s hidden states are upsampled based on the average text-to-unit length ratio, and the resulting features are used to produce speech units, which are then converted into waveforms via a unit-to-speech model. This allows Lyra to generate spoken and text responses simultaneously, leveraging text-derived representations for speech synthesis.

\textbf{(4) MiniCPM-o}, similar to Lyra, is an Omni model that processes vision, speech, and text. For speech, it extends the pre-trained \textit{whisper} encoder by adding a downsampling layer, providing $25$Hz speech features to the LLM. Like other models, the LLM generates text responses from input features, ensuring better semantic coherence than direct speech generation. To enable real-time speech generation, MiniCPM-o employs a streaming speech decoder that takes both the LLM’s hidden features and text response as inputs, generating speech in a chunk-wise autoregressive manner. As a result, MiniCPM-o produces text and speech simultaneously, with speech synthesized in parallel once the number of text tokens reaches a certain chunk size.

\subsubsection{Results on Intermediate Text Responses}
\label{subsubapp:results_on_intermediate_text_responses}
In Section \ref{subsec:retrieval}, we observe that integrating text-based RAG into voice interaction models has minimal impact on overall performance. We confirm that this trend persists in intermediate text responses, demonstrating that errors from speech synthesis do not influence the observed pattern. The results are presented in Figure \ref{fig_rag_results_app}.

\subsubsection{Results on Closed-Source Models}
\label{subsubapp:results_on_closed_source models}
Unlike current closed-source models, which focus on long-context modeling in spoken dialog, we find through extensive experiments that current open-source models struggle with this capability. To establish a performance reference point for long-context modeling in multi-turn scenarios, we evaluate one of the closed-source voice interaction models, \textit{gpt-4o-mini-audio-preview}, on our ContextDialog benchmark. While OpenAI also provides a more advanced version (\textit{gpt-4o-audio-preview}), we used the mini version due to cost constraints.

Currently, OpenAI’s voice interaction API does not support multi-turn spoken interactions where both speakers’ past utterances are explicitly provided as separate audio inputs for use as dialog history. The only available voice history is the actual interaction between the user and the model. Instead, when the past dialog is known in advance, the entire dialog can be provided to the model in one of two ways: (1) providing past utterances as text, with only the current user question and the model’s response in audio (we refer to this setting as \textit{gpt-4o-mini} (text)), or (2) concatenating the multi-turn spoken history with the current user question into a single audio input (\textit{gpt-4o-mini} (audio)).

\begin{table}[t]
\small
\centering
\renewcommand{\arraystretch}{1.2}
\setlength{\tabcolsep}{6pt}
\begin{tabular}{l c c c}
\toprule
\textbf{Model} & \textbf{User} & \textbf{System} & \textbf{Overall} \\
\midrule
\textit{gpt-4o-mini} (text)  & $4.59 \scalebox{0.7}{$\pm 0.04$}$ & $4.42 \scalebox{0.7}{$\pm 0.05$}$ & $4.50 \scalebox{0.7}{$\pm 0.03$}$ \\
\textit{gpt-4o-mini} (audio) & $3.98 \scalebox{0.7}{$\pm 0.07$}$ & $3.68 \scalebox{0.7}{$\pm 0.07$}$ & $3.83 \scalebox{0.7}{$\pm 0.05$}$ \\ \midrule
GLM-4-Voice                        & $1.94 \scalebox{0.7}{$\pm 0.07$}$ & $2.76 \scalebox{0.7}{$\pm 0.08$}$ & $2.35 \scalebox{0.7}{$\pm 0.05$}$ \\
Lyra                               & $2.51 \scalebox{0.7}{$\pm 0.09$}$ & $3.16 \scalebox{0.7}{$\pm 0.09$}$ & $2.83 \scalebox{0.7}{$\pm 0.06$}$ \\
Freeze-Omni                        & $1.73 \scalebox{0.7}{$\pm 0.06$}$ & $2.28 \scalebox{0.7}{$\pm 0.07$}$ & $2.00 \scalebox{0.7}{$\pm 0.05$}$ \\
MiniCPM-o                          & $2.44 \scalebox{0.7}{$\pm 0.09$}$ & $2.84 \scalebox{0.7}{$\pm 0.09$}$ & $2.64 \scalebox{0.7}{$\pm 0.06$}$ \\
\bottomrule
\end{tabular}
\vskip -0.05in
\caption{5-point GPT Score results comparing closed-source and open-source voice interaction models. All results are presented with 95\% confidence intervals. \textit{gpt-4o-mini} (text) refers to the setting where dialog history is provided in text form to \textit{gpt-4o-mini-audio-preview}, and spoken responses are generated via voice interaction. \textit{gpt-4o-mini} (audio) refers to the setting where dialog history is provided in audio form, and spoken responses are used for evaluation.}
\label{tab:closed_source_app}
\vskip -0.2in
\end{table}

\begin{table*}[t]
\small
\centering
\renewcommand{\arraystretch}{1.2} 
\setlength{\tabcolsep}{6pt}
\begin{tabular}{l c c c c c}
\toprule
\multirow{2}{*}{\textbf{Model}} & \multirow{2}{*}{\textbf{LLM FT}} & \multirow{2}{*}{\textbf{Modality}} & \multicolumn{3}{c}{\textbf{Human Evaluation Score}} \\ 
\cmidrule(lr){4-6}
  & & & \textbf{User} & \textbf{System} & \textbf{Overall}  \\ 
\midrule
\rowcolor{gray!10}  &   & $\mathcal{S\rightarrow T,\colorbox{yellow!40}{\hspace{-2pt}$\underline{\bm{\mathcal{S}}}$\hspace{-2pt}}}$ & $2.67 \scalebox{0.7}{$\pm 0.13$}$   & $3.21 \scalebox{0.7}{$\pm 0.13$}$  & $2.93 \scalebox{0.7}{$\pm 0.10$}$  \\ 
\rowcolor{gray!10} \multirow{-2}{*}{GLM-4-Voice \cite{glm2024chatglm}}  & \multirow{-2}{*}{\cmark} & $\mathcal{S\rightarrow \colorbox{yellow!40}{\hspace{-2pt}$\underline{\bm{\mathcal{T}}}$\hspace{-2pt}},S}$ & $2.73 \scalebox{0.7}{$\pm 0.14$}$   & $3.36 \scalebox{0.7}{$\pm 0.13$}$  & $3.04 \scalebox{0.7}{$\pm 0.10$}$ \\ \cdashline{1-6}
glm-4-9b-chat \cite{zeng2024glm4}  &   $-$    & $\mathcal{T\rightarrow \colorbox{yellow!40}{\hspace{-2pt}$\underline{\bm{\mathcal{T}}}$\hspace{-2pt}}}$ & $4.39 \scalebox{0.7}{$\pm 0.10$}$   & $4.27 \scalebox{0.7}{$\pm 0.11$}$  & $4.33 \scalebox{0.7}{$\pm 0.07$}$ \\ 
\midrule
\rowcolor{gray!10}         &   & $\mathcal{S\rightarrow T,\colorbox{yellow!40}{\hspace{-2pt}$\underline{\bm{\mathcal{S}}}$\hspace{-2pt}}}$ & $3.18 \scalebox{0.7}{$\pm 0.12$}$   & $3.39 \scalebox{0.7}{$\pm 0.13$}$  & $3.28 \scalebox{0.7}{$\pm 0.09$}$  \\ 
\rowcolor{gray!10}   \multirow{-2}{*}{Lyra \cite{zhong2024lyra}}        & \multirow{-2}{*}{\cmark}  & $\mathcal{S\rightarrow \colorbox{yellow!40}{\hspace{-2pt}$\underline{\bm{\mathcal{T}}}$\hspace{-2pt}},S}$ & $3.24 \scalebox{0.7}{$\pm 0.12$}$   & $3.51 \scalebox{0.7}{$\pm 0.13$}$  & $3.37 \scalebox{0.7}{$\pm 0.09$}$ \\ \cdashline{1-6}
Qwen2-VL-7B-Instruct \cite{Qwen2-VL}  &    $-$   & $\mathcal{T\rightarrow \colorbox{yellow!40}{\hspace{-2pt}$\underline{\bm{\mathcal{T}}}$\hspace{-2pt}}}$ & $3.76 \scalebox{0.7}{$\pm 0.11$}$   & $3.72 \scalebox{0.7}{$\pm 0.12$}$  & $3.74 \scalebox{0.7}{$\pm 0.08$}$  \\ 
\midrule
\rowcolor{gray!10} &  & $\mathcal{S\rightarrow T,\colorbox{yellow!40}{\hspace{-2pt}$\underline{\bm{\mathcal{S}}}$\hspace{-2pt}}}$ & $2.92 \scalebox{0.7}{$\pm 0.13$}$   & $3.07 \scalebox{0.7}{$\pm 0.12$}$  & $3.00 \scalebox{0.7}{$\pm 0.09$}$ \\ 
\rowcolor{gray!10}  \multirow{-2}{*}{Freeze-Omni \cite{xiong2024freeze}} & \multirow{-2}{*}{\xmark}   & $\mathcal{S\rightarrow \colorbox{yellow!40}{\hspace{-2pt}$\underline{\bm{\mathcal{T}}}$\hspace{-2pt}},S}$ & $3.23 \scalebox{0.7}{$\pm 0.13$}$   & $3.55 \scalebox{0.7}{$\pm 0.12$}$  & $3.39 \scalebox{0.7}{$\pm 0.09$}$  \\ \cdashline{1-6}
Qwen2-7B-Instruct \cite{qwen2}  &   $-$    & $\mathcal{T\rightarrow \colorbox{yellow!40}{\hspace{-2pt}$\underline{\bm{\mathcal{T}}}$\hspace{-2pt}}}$ & $4.08 \scalebox{0.7}{$\pm 0.10$}$   & $3.95 \scalebox{0.7}{$\pm 0.11$}$  & $4.01 \scalebox{0.7}{$\pm 0.08$}$  \\ 
\midrule
\rowcolor{gray!10} &  & $\mathcal{S\rightarrow T,\colorbox{yellow!40}{\hspace{-2pt}$\underline{\bm{\mathcal{S}}}$\hspace{-2pt}}}$ & $2.89 \scalebox{0.7}{$\pm 0.13$}$   & $3.27 \scalebox{0.7}{$\pm 0.12$}$  & $3.09 \scalebox{0.7}{$\pm 0.09$}$  \\ 
\rowcolor{gray!10}    \multirow{-2}{*}{MiniCPM-o \cite{yao2024minicpm}}   & \multirow{-2}{*}{\cmark}   & $\mathcal{S\rightarrow \colorbox{yellow!40}{\hspace{-2pt}$\underline{\bm{\mathcal{T}}}$\hspace{-2pt}},S}$ & $3.42 \scalebox{0.7}{$\pm 0.13$}$   & $3.71 \scalebox{0.7}{$\pm 0.12$}$  & $3.57 \scalebox{0.7}{$\pm 0.09$}$ \\ \cdashline{1-6}
Qwen2.5-7B-Instruct \cite{qwen2.5}  &  $-$     & $\mathcal{T\rightarrow \colorbox{yellow!40}{\hspace{-2pt}$\underline{\bm{\mathcal{T}}}$\hspace{-2pt}}}$ & $4.10 \scalebox{0.7}{$\pm 0.10$}$   & $4.01 \scalebox{0.7}{$\pm 0.10$}$  & $4.05 \scalebox{0.7}{$\pm 0.07$}$  \\ 
\bottomrule
\end{tabular}
\vskip -0.05in
\caption{Human evaluation results for voice interaction models, including the instruct fine-tuned version of each model's backbone LLM. $\mathcal{S}$ and $\mathcal{T}$ represent speech and text, respectively. ``Modality'' indicates input → output data type. ``LLM FT'' shows whether the backbone LLM was fine-tuned during training. ``User'' and ``System'' represent scores for responses to past user and model utterances, respectively. ``Overall'' denotes the score across all responses. All human evaluation results are reported with a 95\% confidence interval.}
\label{tab:recall_human_app}
\end{table*}

As shown in Table \ref{tab:closed_source_app}, the closed-source model, despite receiving only concatenated audio or textual history rather than a true multi-turn voice dialog context, outperforms the open-source models discussed in the main text by a large margin in terms of recall performance. This supports our earlier observation that while some open-source models (e.g., MiniCPM-o) may show competitive performance on certain single-round spoken QA tasks, their ability to handle multi-round context and history remains significantly underdeveloped, even for simple recall tasks. This gap has often been overlooked, and we believe it highlights an important direction for future research.

\subsubsection{Human Evaluation Results}
\label{subsubapp:human evaluation results}

\begin{figure*}[t]
    \centering
    \includegraphics[width=0.98\linewidth]{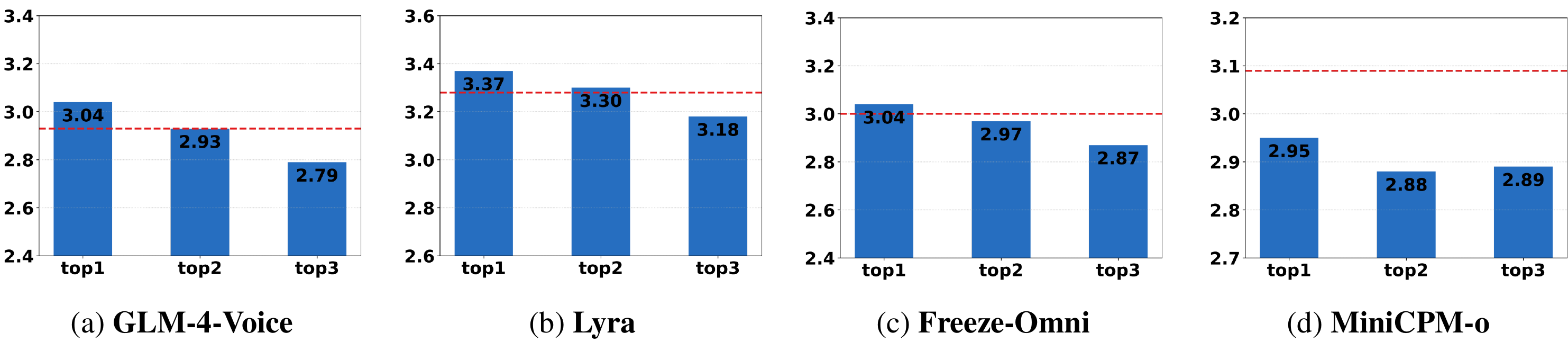}    
    \caption{The human evaluation results ($y$-axis) of applying a RAG method to each model are shown. The red dashed line indicates the results generated without RAG. The evaluation is based on the spoken response $\mathcal{S\rightarrow}\mathcal{T,}\colorbox{yellow!40}{\hspace{-2pt}$\underline{\bm{\mathcal{S}}}$\hspace{-2pt}}$.}
    \label{fig_rag_results_human_app}
    \vskip -0.2in
\end{figure*}

We demonstrate the limitations of the open-source voice interaction model with the GPT Score, compared to its text-based counterpart and to responses generated without RAG, as shown in Table~\ref{tab:recall} and Figure~\ref{fig_rag_results}. In addition, we recruit 180 participants on Amazon Mechanical Turk and conduct a human evaluation. The human evaluation follows the same 5-point scale used for the GPT Score in the main text, measuring how well each model recalls relevant past information. The resulting average scores and their 95\% confidence intervals are reported in Table \ref{tab:recall_human_app} and Figure \ref{fig_rag_results_human_app}.

Although some measurement noise is present, because several participants complete the survey unusually quickly or respond in patterned or apparently random ways, the overall trends remain consistent with those in the main text. Specifically, as shown in Table~\ref{tab:recall_human_app}, the models recall system utterances better than user utterances, and text-based LLMs continue to outperform their speech-based counterparts. Furthermore, as shown in Figure~\ref{fig_rag_results_human_app}, the RAG approach achieves recall performance that is slightly better, similar to, or slightly worse than the baseline.

\subsubsection{Analysis on Additional Models}
\label{subsubapp:additional_models}

We analyze various models in addition to the four models analyzed in the main paper. We provide explanations and results for each model.

\begin{table*}[t]
\small
\centering
\renewcommand{\arraystretch}{1.2} 
\setlength{\tabcolsep}{6pt} 
\begin{tabular}{l c c c c c}
\toprule
\multirow{2}{*}{\textbf{Model}} & \multirow{2}{*}{\textbf{LLM FT}} & \multirow{2}{*}{\textbf{Modality}} & \multicolumn{3}{c}{\textbf{GPT Score}} \\ 
\cmidrule(lr){4-6}
  & & & \textbf{User} & \textbf{System} & \textbf{Overall}  \\ 
\midrule
\rowcolor{gray!10} SLAM-Omni \cite{chen2024slamomnitimbrecontrollablevoiceinteraction}  & \cmark  & $\mathcal{S\rightarrow T,\colorbox{yellow!40}{\hspace{-2pt}$\underline{\bm{\mathcal{S}}}$\hspace{-2pt}}}$ & $1.13\,\scalebox{0.7}{$\pm 0.02$}$   & $1.19\,\scalebox{0.7}{$\pm 0.03$}$  & $1.16\,\scalebox{0.7}{$\pm 0.02$}$ \\ 
Qwen2-0.5B-Instruct \cite{qwen2}  &   $-$    & $\mathcal{T\rightarrow \colorbox{yellow!40}{\hspace{-2pt}$\underline{\bm{\mathcal{T}}}$\hspace{-2pt}}}$ & $1.85\,\scalebox{0.7}{$\pm 0.07$}$   & $1.96\,\scalebox{0.7}{$\pm 0.07$}$  & $1.90\,\scalebox{0.7}{$\pm 0.05$}$  \\ 
\midrule
\rowcolor{gray!10} Moshi \cite{kyutai2024moshi}        & \cmark  & $\mathcal{S\rightarrow T,\colorbox{yellow!40}{\hspace{-2pt}$\underline{\bm{\mathcal{S}}}$\hspace{-2pt}}}$ & $1.16\,\scalebox{0.7}{$\pm 0.03$}$   & $1.55\,\scalebox{0.7}{$\pm 0.06$}$  & $1.35\,\scalebox{0.7}{$\pm 0.03$}$ \\ 
\bottomrule
\end{tabular}
\caption{Evaluation results for additional open-source multi-turn voice interaction models, including the instruct-tuned versions of their backbone LLMs.
$\mathcal{S}$ represents speech, and $\mathcal{T}$ represents text. ``Modality'' denotes the input $\rightarrow$ output data type for each model, while ``LLM FT'' indicates whether the backbone LLM is fine-tuned or kept frozen during training. ``User'' refers to scores for responses to questions about past user utterances, whereas ``System'' assesses responses regarding the model’s own past utterances. ``Overall'' represents the average score across all responses. Scores are reported with a 95\% confidence interval.}
\label{tab:additional_recall}
\vskip -0.2in
\end{table*}

\textbf{Single-Round Voice Interaction Models} Several models, including SpeechGPT \cite{zhang-etal-2023-speechgpt}, USDM \cite{kim2024paralinguisticsaware}, LLaMa-Omni \cite{fang2025llamaomni}, and Mini-Omni \cite{xie2024miniomnilanguagemodelshear}, are trained on single-round data. While some official implementations support multi-turn settings, they do not retain conversation history, treating each exchange as an independent query-response pair. When we modified these models to incorporate conversation history, discrepancies between training and inference led to unreliable multi-turn generation. Due to these limitations, we exclude them from our main analysis.

\textbf{Multi-Round Voice Interaction Models} We evaluate the recall capabilities of the open-source voice interaction models, Moshi \cite{kyutai2024moshi} and SLAM-Omni \cite{chen2024slamomnitimbrecontrollablevoiceinteraction}, using the same methodology as in Section \ref{subsec:recall}. A brief description of each model follows.

\textbf{(1) Moshi} is a voice interaction model built using Mimi, a streaming neural audio codec trained with residual vector quantization \cite{9625818}. Mimi extracts an 8-level codec at $12.5$Hz from both input and output speech during training. The core interaction model, trained on these codec features together with text, models speech from both the user and the system. Unlike the four previously analyzed models, which handle only one speaker at a time, Moshi enables flexible interactions (e.g., backchanneling, interruptions) by jointly modeling an 8-level user codec and an 8-level system codec, resulting in 16 tokens per time step. Additionally, to prevent semantic degradation, Moshi generates response text tokens along with 16 speech tokens, producing a total of 17 tokens per time step.

Since response text tokens ($3\sim4$Hz) are significantly shorter than speech tokens ($12.5$Hz), Moshi employs speech-aligned text tokens ($12.5$Hz), leveraging pre-extracted text-speech alignment during training. However, for RAG-based analysis in Section \ref{subsec:retrieval}, the prompt provided to Moshi must also be an expanded sequence aligned with speech, similar to training, which cannot be derived from text alone. Due to this limitation, we conduct only the recall analysis from Section \ref{subsec:recall} for Moshi.

\textbf{(2) SLAM-Omni} is another model that processes speech input and generates speech output. The input speech is encoded using \textit{whisper}, and the extracted features pass through a projector that aligns embeddings before being fed into the interaction model. The model produces discrete semantic tokens at $50$Hz, following the approach used in CosyVoice-300M-SFT \cite{du2024cosyvoicescalablemultilingualzeroshot}. To mitigate the challenges of storing past conversations as speech, which would significantly increase length and degrade long-context performance, SLAM-Omni retains all past interactions as text. It utilizes text dialog history along with the user's current speech input to generate responses. We evaluate SLAM-Omni’s recall performance using the same methodology as in Section \ref{subsec:recall}.

\begin{table*}[t]
\small
\centering
\renewcommand{\arraystretch}{1.2} 
\setlength{\tabcolsep}{6pt}
\begin{tabular}{l c c c c c c}
\toprule
\multirow{2}{*}{\textbf{Model}} & \multirow{2}{*}{\textbf{LLM FT}} & \multirow{2}{*}{\textbf{Modality}} & \multicolumn{3}{c}{\textbf{GPT Score}} & \multirow{2}{*}{\textbf{WER}} \\ 
\cmidrule(lr){4-6}
  & & & \textbf{User} & \textbf{System} & \textbf{Overall} & \\ 
\midrule
\rowcolor{gray!10}  &   & $\mathcal{S\rightarrow T,\colorbox{yellow!40}{\hspace{-2pt}$\underline{\bm{\mathcal{S}}}$\hspace{-2pt}}}$ & $1.43 \scalebox{0.7}{$\pm 0.07$}$   & $2.25 \scalebox{0.7}{$\pm 0.11$}$  & $1.84 \scalebox{0.7}{$\pm 0.07$}$  & 15.72\% \\ 
\rowcolor{gray!10} \multirow{-2}{*}{GLM-4-Voice \cite{glm2024chatglm}}  & \multirow{-2}{*}{\cmark} & $\mathcal{S\rightarrow \colorbox{yellow!40}{\hspace{-2pt}$\underline{\bm{\mathcal{T}}}$\hspace{-2pt}},S}$ & $1.55 \scalebox{0.7}{$\pm 0.08$}$   & $2.89 \scalebox{0.7}{$\pm 0.12$}$  & $2.22 \scalebox{0.7}{$\pm 0.08$}$  & $-$ \\ \cdashline{1-7} 
glm-4-9b-chat \cite{zeng2024glm4}  &   $-$    & $\mathcal{T\rightarrow \colorbox{yellow!40}{\hspace{-2pt}$\underline{\bm{\mathcal{T}}}$\hspace{-2pt}}}$ & $4.06 \scalebox{0.7}{$\pm 0.07$}$   & $4.49 \scalebox{0.7}{$\pm 0.06$}$  & $4.28 \scalebox{0.7}{$\pm 0.05$}$  & $-$ \\ 
\midrule
\rowcolor{gray!10}         &   & $\mathcal{S\rightarrow T,\colorbox{yellow!40}{\hspace{-2pt}$\underline{\bm{\mathcal{S}}}$\hspace{-2pt}}}$ & $2.66 \scalebox{0.7}{$\pm 0.11$}$   & $3.37 \scalebox{0.7}{$\pm 0.11$}$  & $3.02 \scalebox{0.7}{$\pm 0.08$}$  & 34.66\% \\ 
\rowcolor{gray!10}   \multirow{-2}{*}{Lyra \cite{zhong2024lyra}}        & \multirow{-2}{*}{\cmark}  & $\mathcal{S\rightarrow \colorbox{yellow!40}{\hspace{-2pt}$\underline{\bm{\mathcal{T}}}$\hspace{-2pt}},S}$ & $2.86 \scalebox{0.7}{$\pm 0.11$}$   & $4.20 \scalebox{0.7}{$\pm 0.09$}$  & $3.53 \scalebox{0.7}{$\pm 0.08$}$  & $-$ \\ \cdashline{1-7} 
Qwen2-VL-7B-Instruct \cite{Qwen2-VL}  &    $-$   & $\mathcal{T\rightarrow \colorbox{yellow!40}{\hspace{-2pt}$\underline{\bm{\mathcal{T}}}$\hspace{-2pt}}}$ & $4.02 \scalebox{0.7}{$\pm 0.09$}$   & $4.16 \scalebox{0.7}{$\pm 0.10$}$  & $4.09 \scalebox{0.7}{$\pm 0.07$}$  & $-$ \\ 
\midrule
\rowcolor{gray!10} &  & $\mathcal{S\rightarrow T,\colorbox{yellow!40}{\hspace{-2pt}$\underline{\bm{\mathcal{S}}}$\hspace{-2pt}}}$ & $2.31 \scalebox{0.7}{$\pm 0.10$}$   & $2.45 \scalebox{0.7}{$\pm 0.10$}$  & $2.38 \scalebox{0.7}{$\pm 0.07$}$  & 12.37\% \\ 
\rowcolor{gray!10}  \multirow{-2}{*}{Freeze-Omni \cite{xiong2024freeze}} & \multirow{-2}{*}{\xmark}   & $\mathcal{S\rightarrow \colorbox{yellow!40}{\hspace{-2pt}$\underline{\bm{\mathcal{T}}}$\hspace{-2pt}},S}$ & $2.67 \scalebox{0.7}{$\pm 0.11$}$   & $3.73 \scalebox{0.7}{$\pm 0.11$}$  & $3.20 \scalebox{0.7}{$\pm 0.08$}$  & $-$ \\ \cdashline{1-7} 
Qwen2-7B-Instruct \cite{qwen2}  &   $-$    & $\mathcal{T\rightarrow \colorbox{yellow!40}{\hspace{-2pt}$\underline{\bm{\mathcal{T}}}$\hspace{-2pt}}}$ & $4.23 \scalebox{0.7}{$\pm 0.07$}$   & $4.49 \scalebox{0.7}{$\pm 0.07$}$  & $4.36 \scalebox{0.7}{$\pm 0.05$}$  & $-$ \\ 
\midrule
\rowcolor{gray!10} &  & $\mathcal{S\rightarrow T,\colorbox{yellow!40}{\hspace{-2pt}$\underline{\bm{\mathcal{S}}}$\hspace{-2pt}}}$ & $2.01 \scalebox{0.7}{$\pm 0.10$}$   & $1.61 \scalebox{0.7}{$\pm 0.08$}$  & $1.81 \scalebox{0.7}{$\pm 0.06$}$  & 71.03\% \\ 
\rowcolor{gray!10}    \multirow{-2}{*}{MiniCPM-o \cite{yao2024minicpm}}   & \multirow{-2}{*}{\cmark}   & $\mathcal{S\rightarrow \colorbox{yellow!40}{\hspace{-2pt}$\underline{\bm{\mathcal{T}}}$\hspace{-2pt}},S}$ & $3.36\scalebox{0.7}{$\pm 0.11$}$   & $4.25 \scalebox{0.7}{$\pm 0.09$}$  & $3.81 \scalebox{0.7}{$\pm 0.07$}$  & $-$ \\ \cdashline{1-7} 
Qwen2.5-7B-Instruct \cite{qwen2.5}  &  $-$     & $\mathcal{T\rightarrow \colorbox{yellow!40}{\hspace{-2pt}$\underline{\bm{\mathcal{T}}}$\hspace{-2pt}}}$ & $4.09 \scalebox{0.7}{$\pm 0.07$}$   & $4.39 \scalebox{0.7}{$\pm 0.07$}$  & $4.24 \scalebox{0.7}{$\pm 0.05$}$  & $-$ \\ 
\bottomrule
\end{tabular}
\vskip -0.05in
\caption{Evaluation results for the additional dataset constructed using the SpokenWOZ dataset. The definitions of each term and the evaluation method follow those in Table \ref{tab:recall}.}
\label{tab:recall_woz}
\vskip -0.05in
\end{table*}

\begin{figure*}[h]
    \centering
    \includegraphics[width=0.95\linewidth]{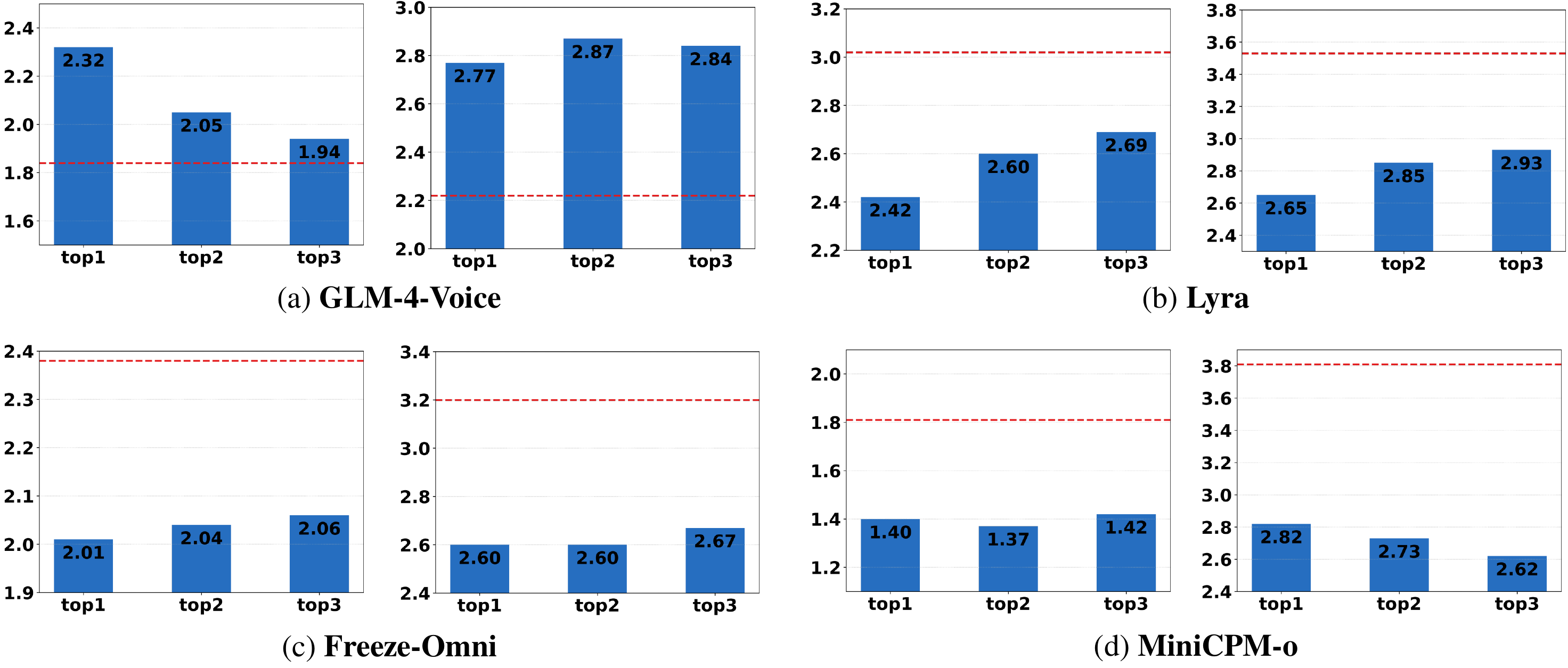}
    \vskip -0.05in
    \caption{The results of applying a RAG method to each model are shown, where the left side of each subfigure represents the evaluation on the spoken response ($\mathcal{S\rightarrow T,\colorbox{yellow!40}{\hspace{-2pt}$\underline{\bm{\mathcal{S}}}$\hspace{-2pt}}}$), and the right side represents the evaluation on the intermediate text response ($\mathcal{S\rightarrow \colorbox{yellow!40}{\hspace{-2pt}$\underline{\bm{\mathcal{T}}}$\hspace{-2pt}},S}$). The red dashed line indicates the results generated without RAG.}
    \label{fig_rag_results_woz}
    \vskip -0.22in
\end{figure*}

\begin{table*}[t]
\small
\centering
\renewcommand{\arraystretch}{1.2} 
\setlength{\tabcolsep}{6pt}
\begin{tabular}{l l c c c}
\toprule
\textbf{Model} & \textbf{Prompt} & \textbf{top-\textit{1}} & \textbf{top-\textit{2}} & \textbf{top-\textit{3}} \\ 
\midrule
\multirow{4}{*}{\centering GLM-4-Voice \cite{glm2024chatglm}} 
 & \texttt{Based on your/my statement ...} & $2.34 \scalebox{0.7}{$\pm 0.05$}$ & $2.30 \scalebox{0.7}{$\pm 0.05$}$ & $2.09 \scalebox{0.7}{$\pm 0.05$}$ \\ 
 & \texttt{Since you/I said ...}              & $2.09 \scalebox{0.7}{$\pm 0.05$}$ & $1.93 \scalebox{0.7}{$\pm 0.05$}$ & $1.52 \scalebox{0.7}{$\pm 0.04$}$ \\ 
 & \texttt{As I recall you/myself saying ...}   & $1.60 \scalebox{0.7}{$\pm 0.04$}$ & $1.49 \scalebox{0.7}{$\pm 0.04$}$ & $1.39 \scalebox{0.7}{$\pm 0.04$}$ \\ 
 & Concatenation of Utterances              & $1.55 \scalebox{0.7}{$\pm 0.04$}$ & $1.47 \scalebox{0.7}{$\pm 0.04$}$ & $1.35 \scalebox{0.7}{$\pm 0.04$}$ \\ 
\midrule
\multirow{4}{*}{\centering Lyra \cite{zhong2024lyra}}
 & \texttt{Based on your/my statement ...} & $2.83 \scalebox{0.7}{$\pm 0.06$}$ & $2.68 \scalebox{0.7}{$\pm 0.06$}$ & $2.52 \scalebox{0.7}{$\pm 0.06$}$ \\ 
 & \texttt{Since you/I said ...}              & $2.02 \scalebox{0.7}{$\pm 0.05$}$ & $1.98 \scalebox{0.7}{$\pm 0.05$}$ & $1.60 \scalebox{0.7}{$\pm 0.05$}$ \\ 
 & \texttt{As I recall you/myself saying ...}   & $1.57 \scalebox{0.7}{$\pm 0.05$}$ & $1.42 \scalebox{0.7}{$\pm 0.04$}$ & $1.56 \scalebox{0.7}{$\pm 0.04$}$ \\ 
 & Concatenation of Utterances              & $1.71 \scalebox{0.7}{$\pm 0.05$}$ & $1.52 \scalebox{0.7}{$\pm 0.04$}$ & $1.46 \scalebox{0.7}{$\pm 0.04$}$ \\ 
\midrule
\multirow{4}{*}{\centering Freeze-Omni \cite{xiong2024freeze}}
 & \texttt{Based on your/my statement ...} & $2.02 \scalebox{0.7}{$\pm 0.04$}$ & $1.98 \scalebox{0.7}{$\pm 0.04$}$ & $1.80 \scalebox{0.7}{$\pm 0.04$}$ \\ 
 & \texttt{Since you/I said ...}              & $2.00 \scalebox{0.7}{$\pm 0.04$}$ & $1.95 \scalebox{0.7}{$\pm 0.04$}$ & $1.64 \scalebox{0.7}{$\pm 0.04$}$ \\ 
 & \texttt{As I recall you/myself saying ...}   & $1.98 \scalebox{0.7}{$\pm 0.04$}$ & $1.76 \scalebox{0.7}{$\pm 0.04$}$ & $1.66 \scalebox{0.7}{$\pm 0.04$}$ \\ 
 & Concatenation of Utterances              & $1.40 \scalebox{0.7}{$\pm 0.03$}$ & $1.22 \scalebox{0.7}{$\pm 0.03$}$ & $1.19 \scalebox{0.7}{$\pm 0.03$}$ \\ 
\midrule
\multirow{4}{*}{\centering MiniCPM-o \cite{yao2024minicpm}}
 & \texttt{Based on your/my statement ...} & $2.10 \scalebox{0.7}{$\pm 0.05$}$ & $1.91 \scalebox{0.7}{$\pm 0.05$}$ & $1.81 \scalebox{0.7}{$\pm 0.05$}$ \\ 
 & \texttt{Since you/I said ...}              & $1.67 \scalebox{0.7}{$\pm 0.05$}$ & $1.57 \scalebox{0.7}{$\pm 0.04$}$ & $1.37 \scalebox{0.7}{$\pm 0.04$}$ \\ 
 & \texttt{As I recall you/myself saying ...}   & $1.54 \scalebox{0.7}{$\pm 0.04$}$ & $1.44 \scalebox{0.7}{$\pm 0.04$}$ & $1.50 \scalebox{0.7}{$\pm 0.04$}$ \\ 
 & Concatenation of Utterances              & $1.39 \scalebox{0.7}{$\pm 0.04$}$ & $1.28 \scalebox{0.7}{$\pm 0.03$}$ & $1.18 \scalebox{0.7}{$\pm 0.03$}$ \\ 
\bottomrule
\end{tabular}
\vskip -0.05in
\caption{Evaluation results for the effects of prompts used in RAG. For each model, the same four prompts are evaluated, with GPT Scores reported along with a 95\% confidence interval.}
\label{tab:recall_rag_prompt}
\vskip -0.2in
\end{table*}

The recall performance of past utterances for both models is in Table \ref{tab:additional_recall}. We reaffirm two key findings from Section \ref{subsec:recall}: (1) SLAM-Omni performs worse than its text-based counterparts (excluding Moshi, as its backbone LLM is unavailable), and (2) Moshi, which processes user inputs solely through speech, shows significantly lower recall performance for user utterances than for model-generated ones ($p<0.01$). SLAM-Omni is excluded from this comparison as it retains past interactions in text format for both user and model. These results further validate the generalizability of our analysis. Notably, both models exhibit lower performance than those analyzed in the main paper.

While multiple factors contribute to performance drops, SLAM-Omni’s small backbone model size is the most likely cause. Unlike other models with 7B$\sim$9B parameters, SLAM-Omni is built on a much smaller 0.5B LLM, and even its chat variant exhibits weak recall performance.

For Moshi, multiple factors may contribute to its performance degradation. Unlike models with clearly separated input and output, Moshi processes both speakers’ voices simultaneously, allowing flexible interactions (e.g., interruptions) without strict turn-taking. Consequently, it sometimes remains silent instead of responding to past conversations and user queries, leading to performance loss. Additionally, as a free-form conversational model, Moshi lacks explicit end markers for speech output, making it difficult to determine when to stop generation. To ensure a consistent evaluation, we assess speech generated within a fixed 12-second window, though this may introduce artifacts such as unintended utterances or truncated responses, further impacting performance.

\subsubsection{Analysis on Additional Dataset}
\label{subsubapp:additional_datasets}
To further enhance the reliability of our analysis, we create an additional dataset following a similar pipeline described in Section \ref{sec:dataset} to evaluate both the recall ability and RAG performance of voice interaction models. For this dataset, we use spoken dialog data from SpokenWOZ \cite{NEURIPS2023_7b16688a}, a task-oriented dialog dataset where users interact with the model to achieve specific goals such as booking flights or making restaurant reservations. This dataset closely resembles real-world voice assistant applications, where remembering past user utterances is crucial.

We construct a new Spoken QA dataset using the \texttt{test} split of SpokenWOZ. Using 1,000 dialogs spanning approximately 44 hours, we create 1,930 QA pairs, with each dialog requiring the recall of both user and model utterances, along with their corresponding supporting utterances. Compared to MultiDialog, which has an average conversation length of around 2.5 minutes, SpokenWOZ consists of longer conversations averaging 6.5 minutes. This allows us to evaluate model recall performance over extended dialogs and assess the impact of augmenting retrieved sentences during generation.

However, the SpokenWOZ transcripts are generated using an ASR model and contain discrepancies from the original audio, meaning the QA data derived from them may not be perfectly aligned with the original spoken dialog. Additionally, the original audio quality is low at 8kHz, making it unsuitable for high-fidelity analysis. Therefore, we include the results from this dataset as a reference in the Appendix. The recall performance of each voice interaction model is presented in Table \ref{tab:recall_woz}, while the spoken response performance with retrieved sentences from the dedicated module, \textit{e5-large-v2} \cite{Wang2022TextEB}, is in Figure \ref{fig_rag_results_woz}.

\begin{table*}[t]
\small
\begin{center}
\begin{tabular}{l|l|l}
\toprule
\textbf{Type} & \textbf{Models}                                                         & \textbf{Link}                                                                                                                 \\ \midrule
\multirow{6}{*}{\shortstack[c]{Voice}} 
                 & GLM-4-Voice \cite{glm2024chatglm}                                       & \href{https://github.com/THUDM/GLM-4-Voice}{https://github.com/THUDM/GLM-4-Voice}                                             \\ 
                 & Freeze-Omni \cite{xiong2024freeze}                                      & \href{https://github.com/VITA-MLLM/Freeze-Omni}{https://github.com/VITA-MLLM/Freeze-Omni}                                     \\ 
                 & Lyra \cite{zhong2024lyra}                                               & \href{https://github.com/dvlab-research/Lyra}{https://github.com/dvlab-research/Lyra}                                         \\ 
                 & MiniCPM-o \cite{yao2024minicpm}                                         & \href{https://github.com/OpenBMB/MiniCPM-o}{https://github.com/OpenBMB/MiniCPM-o}                                             \\ 
                 & Slam-Omni \cite{chen2024slamomnitimbrecontrollablevoiceinteraction}     & \href{https://github.com/X-LANCE/SLAM-LLM}{https://github.com/X-LANCE/SLAM-LLM}                                               \\ 
                 & Moshi \cite{kyutai2024moshi}                                            & \href{https://github.com/kyutai-labs/moshi}{https://github.com/kyutai-labs/moshi}                                             \\ \midrule
\multirow{5}{*}{\shortstack[c]{Text}} 
                 & glm-4-9b-chat \cite{zeng2024glm4}                                       & \href{https://huggingface.co/THUDM/glm-4-9b-chat}{https://huggingface.co/THUDM/glm-4-9b-chat}                                 \\ 
                 & Qwen2-VL-7B-Instruct \cite{qwen2}                                       & \href{https://huggingface.co/Qwen/Qwen2-VL-7B-Instruct}{https://huggingface.co/Qwen/Qwen2-VL-7B-Instruct}                     \\ 
                 & Qwen2-7B-Instruct \cite{Qwen2-VL}                                       & \href{https://huggingface.co/Qwen/Qwen2-7B-Instruct}{https://huggingface.co/Qwen/Qwen2-7B-Instruct}                           \\ 
                 & Qwen2.5-7B-Instruct \cite{qwen2.5}                                      & \href{https://huggingface.co/Qwen/Qwen2.5-7B-Instruct}{https://huggingface.co/Qwen/Qwen2.5-7B-Instruct}                       \\ 
                 & Qwen2-0.5B-Instruct \cite{qwen2}                                        & \href{https://huggingface.co/Qwen/Qwen2-0.5B-Instruct}{https://huggingface.co/Qwen/Qwen2-0.5B-Instruct}                       \\ \midrule
\multirow{9}{*}{\shortstack[c]{Extra}} 
                 & Fish Speech \cite{fish-speech-v1.4}                                     & \href{https://github.com/fishaudio/fish-speech}{https://github.com/fishaudio/fish-speech}                                     \\ 
                 & \textit{whisper-large-v3} \cite{pmlr-v202-radford23a}                   & \href{https://huggingface.co/openai/whisper-large-v3}{https://huggingface.co/openai/whisper-large-v3}                         \\ 
                 & \textit{whisper-large-v3-turbo} \cite{pmlr-v202-radford23a}             & \href{https://huggingface.co/openai/whisper-large-v3-turbo}{https://huggingface.co/openai/whisper-large-v3-turbo}             \\ 
                 & \textit{e5-large-v2} \cite{Wang2022TextEB}                              & \href{https://huggingface.co/intfloat/e5-large-v2}{https://huggingface.co/intfloat/e5-large-v2}                               \\ 
                 & SONAR \cite{Duquenne:2023:sonar_arxiv}                                  & \href{https://github.com/facebookresearch/SONAR}{https://github.com/facebookresearch/SONAR}                                   \\
                 & \textit{gpt-4o (24-08-06)} \cite{openai2024gpt4ocard}                 & \href{https://platform.openai.com/docs/models#gpt-4o}{https://platform.openai.com/docs/models\#gpt-4o}                        \\ 
                 & \textit{gpt-4o-mini (24-07-18)} \cite{openai2024gpt4ocard}            & \href{https://platform.openai.com/docs/models#gpt-4o-mini}{https://platform.openai.com/docs/models\#gpt-4o-mini}               \\ 
                 & \textit{gpt-4o-mini-audio-preview (24-12-17)} \cite{openai2024gpt4ocard}
& \makecell[l]{\href{https://platform.openai.com/docs/models/gpt-4o-mini-audio-preview}%
{https://platform.openai.com/docs/models}\\
\href{https://platform.openai.com/docs/models/gpt-4o-mini-audio-preview}{\#gpt-4o-mini-audio-preview}} \\
                 & \textit{o1-mini (24-07-18)} \cite{openai2024openaio1card}             & \href{https://platform.openai.com/docs/models#o1}{https://platform.openai.com/docs/models\#o1}                                \\   
                 & \textit{\texttt{evaluate}} \cite{von-werra-etal-2022-evaluate}          & \href{https://github.com/huggingface/evaluate}{https://github.com/huggingface/evaluate}                                       \\   
                 \bottomrule
\end{tabular}
\vskip -0.05in
\caption{Links to the models, libraries, APIs, and checkpoints used in our experiments.}
\vskip -0.1in
\label{table:model_links}
\end{center}
\end{table*}

\begin{table*}[t]
\small
\begin{center}
\begin{tabular}{l|l|cc}
\toprule
\textbf{Type} & \textbf{Name}                  & \textbf{Speech}       & \textbf{License}     \\ \midrule
\multirow{8}{*}{Model} 
& GLM-4-Voice \cite{glm2024chatglm}       & \cmark                & Apache-2.0 \\
& glm-4-9b-chat \cite{zeng2024glm4}       & \xmark                & Apache-2.0 \\
& Lyra \cite{zhong2024lyra}               & \cmark                & Apache-2.0 \\
& Qwen2-VL-7B-Instruct \cite{Qwen2-VL}    & \xmark                & Apache-2.0 \\
& Freeze-Omni \cite{xiong2024freeze}      & \cmark                & Apache-2.0 \\
& Qwen2-7B-Instruct \cite{qwen2}          & \xmark                & Apache-2.0 \\
& MiniCPM-o \cite{yao2024minicpm}         & \cmark                & Apache-2.0 \\
& Qwen2.5-7B-Instruct \cite{qwen2.5}      & \xmark                & Apache-2.0 \\ 
& SLAM-Omni \cite{chen2024slamomnitimbrecontrollablevoiceinteraction}        & \cmark                & MIT License \\
& Qwen2-0.5B-Instruct \cite{qwen2}      & \xmark                & Apache-2.0 \\ 
& Moshi \cite{kyutai2024moshi}         & \cmark                & Apache-2.0, MIT License \\ \midrule
\multirow{3}{*}{Dataset} 
& MultiDialog \cite{park-etal-2024-lets}  & \cmark                & CC \\
& SpokenWoz \cite{NEURIPS2023_7b16688a}   & \cmark                & CC BY-NC 4.0 \\
& ContextDialog                           & \cmark                & CC BY-NC 4.0 \\
\bottomrule
\end{tabular}
\vskip -0.05in
\caption{License and relevance to speech of each model and dataset used for analyses.}
\vskip -0.2in
\label{table:model_license}
\end{center}
\end{table*}

Overall, the trends remain consistent with the main paper: (1) performance degradation compared to text-based counterparts, particularly in recalling past user utterances, and (2) minimal impact of RAG on improving past information-based QA accuracy. However, MiniCPM-o and GLM-4-Voice exhibit opposite trends in the two respective experiments compared to the original findings. Given the quality issues in SpokenWOZ, as also evidenced by the high WER in Table \ref{tab:recall_woz}, we emphasize that these results are for reference only.

\subsubsection{Analysis on Retrieval Prompts}
\label{subsubapp:retrieval_prompt}
In this section, we assess whether the observation from Figure \ref{fig_rag_results} in Section \ref{subsec:retrieval}—that RAG generally has limited effectiveness in voice interaction models—holds across different prompt templates beyond the \texttt{Based on your/my statement: ...} format used in the main paper. We conduct experiments using three additional prompt templates: (1) \texttt{Since you/I said ...}, (2) \texttt{As I recall you/myself saying ...}, and (3) a simple concatenation of retrieved sentences. The evaluation is performed on the final spoken response, and the results are presented in Table \ref{tab:recall_rag_prompt}.

From Table \ref{tab:recall_rag_prompt}, we observe that the prompt used in our main experiments minimizes performance degradation compared to other prompts, which also exhibit similar declines. Additionally, performance varies significantly depending on the prompt used for RAG. Considering these results with Table \ref{tab:rag_support}, which demonstrates performance improvements when providing supporting utterances, our findings indicate that the prompt template we used is effective for RAG. However, further exploration of more optimal prompting and augmentation strategies tailored for spoken response generation in voice interaction models remains a key research direction.

\subsubsection{Failure Cases}
\label{subsubapp:failure_cases}

In this section, we categorize the responses from models that received low scores in our evaluation. Figure \ref{app:failure_cases} illustrates common types of errors these models frequently make.

(1) The first type of error occurs when the model expresses uncertainty, stating that it does not recall the necessary information, likely due to its failure to retrieve past utterances. (2) The second type involves retrieving an incorrect utterance, leading to an erroneous answer. As shown in Example 2, to correctly respond to the question, the model should refer to an utterance related to a sitcom; however, it mistakenly retrieves an unrelated one, resulting in a wrong response.

(3) The third type involves generating an incorrect answer by relying on intrinsic knowledge rather than recalling the relevant past utterance. For instance, even though the necessary information was mentioned earlier in the conversation, an unrelated topic—the Lion King—appeared later, causing the model to mistakenly respond with Lion King. Notably, while the conversation never mentioned that The Lion King was released in 1994, the model included this fact based solely on its intrinsic knowledge.

(4) Finally, some cases exhibit multiple error types simultaneously. In Example 4, even though ``Oklahoma!'' was never mentioned in the conversation, the model generated a response including this term. Upon investigation, we found that while ``Oklahoma!' is the title of a state anthem, it is unrelated to Jimmy Rogers and has no connection to the year 1971, highlighting the model’s tendency to produce hallucinated responses.

\subsection{Licenses and Links}
\label{subapp:licenses}
The links and licenses for the models, datasets, and libraries used in our experiments and analyses, along with ContextDialog, are listed in Table \ref{table:model_links} and Table \ref{table:model_license}, respectively. Our benchmark is intended solely for research on voice interaction models. For writing, we use ChatGPT-4o exclusively for expression and grammar refinement.

\begin{figure*}[ht]
\begin{center}
\begin{tcolorbox}[colback=white, coltext=black, title=\textbf{Failure Cases}]

\textbf{[Example 1] I Don't Know}

\ldots

User: Yeah my nephews watch Pokemon on Netflix, I found it too silly though. I did not know the UK did their own version of thats 70s show that was canceled after 10 episodes though.

\textcolor{blue}{System: yeah, that show was ok, kinda funny from what I remember. Hard to believe only 1 episode of the simpsons without bart, mentioned or appearing.}

User: Yeah I still watch it on Netflix every now and then if I am bored, I have been re watching game of thrones as of late though.

\ldots

User: Could you remind me what you mentioned about the episode of The Simpsons without Bart being mentioned or appearing?

\textcolor{red}{System: I don't remember that.}

\bigskip

\textbf{[Example 2] Wrong Evidence}

\ldots

\textcolor{blue}{System: My favoite sitcom was friends. I miss it. I did not think 3rd rock from the sun deserved the golden globe in 1997 over friends.}

User: Friends was a great show, they had so much talent there. I was and am more of a Seinfeld fan, 3rd rock was ok.

System: I just love that comedies generally have a happy ending. \textbf{I love to smile and laugh}.

User: Yeah same here, and comedy is pretty hard to do right!

\ldots

User: Could you remind me what you said about your favorite sitcom and its recognition in award shows?

\textcolor{red}{System: I love that comedies generally have a happy ending. I love to smile and laugh.}

\bigskip

\textbf{[Example 3] Incorrect Answer with Instrinsic Knowledge}

User: Did you watch Infinity Wars?

\textcolor{blue}{System: No. I heard that it was the first film ever shot in digital imax.}

\ldots

System: That is neat! How about \textbf{"The Lion King"} being dubbed into Zulu for its African release? That is also neat.

\ldots

User: Could you remind me what you mentioned about the first film shot in digital IMAX?

\textcolor{red}{System: The first film shot in digital IMAX was \textbf{"The Lion King"}. It was released in \textbf{1994}.}

\bigskip

\textbf{[Example 4] Wrong Evidence with Incorrect Knowledge}

\textcolor{blue}{User: That's too funny. Did you know that the state rock song of Oklahoma is "do you realize?" by the Flaming Lips?}

System: I didn't know states could have state rock songs! I wonder if Illinois' is that song that became the most requested track on that Chicago alt-rock radio station.

\ldots

User: Could you remind me of which song I mentioned as the state rock song of Oklahoma?

\textcolor{red}{System: Yes, I believe you're referring to \textbf{"Oklahoma!"} which is the state rock song of Oklahoma. It was written by \textbf{Jimmy Rogers} and was adopted as the state song in \textbf{1971}.}

\end{tcolorbox}
\end{center}
\caption{Examples categorizing common model failure cases. This excerpt is from a conversation: \textbf{\textcolor{blue}{blue}} text indicates the supporting utterance, while \textbf{\textcolor{red}{red}} text highlights the incorrect response.}
\label{app:failure_cases}
\end{figure*}

\subsection{ContextDialog}
\label{subapp:contextdialog}
In this section, we detail the customized prompt design used to construct ContextDialog and provide examples of the generated data.

\subsubsection{Prompt for ContextDialog}
\label{subsubapp:templates}

\textbf{Generation Prompt}
We use \textit{gpt-4o} \cite{openai2024gpt4ocard} as a QA generator, applying a custom-designed prompt to generate text-based questions, answers, and supporting utterances from dialog transcripts. As described in Section \ref{subsec:generation}, ContextDialog contains four QA pairs per spoken dialog, determined by the placement and speaker identity of the supporting utterance. To construct these pairs, we reuse the generation prompt with minimal modifications (e.g., replacing ``first half'' with ``latter half'' or ``system said'' with ``user said''). Figure \ref{app:generation-prompt-2} shows an example prompt used to generate question, answer, and supporting utterance pairs based on a \textit{system utterance} from the \textit{first half} of the conversation.

When designing this prompt, we consider several key factors. First, our goal is to simulate real-world scenarios where people forget and reconfirm information. To achieve this, we structure each question to double-check a single relevant utterance from either the user or the system. The prompt ensures that the user’s question and the system’s response naturally relate to the preceding dialog (Requirements 4, 7, and 9). Additionally, since we aim to evaluate voice interaction models in realistic settings, we prioritize detailed answers over simple yes/no responses (Requirement 8).

To enhance benchmark completeness and usability, we enforce specific requirements. Requirement 1 ensures that questions are generated solely from information that appears only once in the conversation. This constraint prevents confusion caused by participants correcting themselves or changing decisions mid-dialog; otherwise, a QA pair might seem valid when considering only the supporting utterance but become misleading when viewed in full context. Moreover, Requirement 3 mandates that the supporting utterance be provided alongside the generated QA pair. This metadata serves as a precise reference for dialog history and is essential for evaluating augmented generation (Section \ref{subsec:retrieval}).

\textbf{Validation Prompt}
For validation, we use \textit{o1-mini} \cite{openai2024gpt4ocard} as a reviewer, applying a customized validation prompt—shown in Figure \ref{app:validation-prompt}—to assess the generated question, answer, and supporting utterance pairs. This prompt ensures that the answer is fully deducible when a portion of the dialog history is provided alongside the generated QA pair. Following the validation process described in Section \ref{subsec:generation}, we obtain QA pairs in written form that meet our predefined criteria.

\textbf{Refining Prompt for Text-to-Speech}
To convert the validated text-based QA pairs into speech using the voices of both speakers in the conversation, we use Fish Speech \cite{fish-speech-v1.4}, a speaker-adaptive TTS model that synthesizes speech in the target speaker’s timbre using reference audio. Before synthesis, we normalize the text QA data into a TTS-compatible format using the refine prompt in Figure \ref{app:refine-prompt-2} with \textit{gpt-4o}.

\begin{figure*}[ht]
\begin{center}
\begin{tcolorbox}[colback=white, coltext=black, title=\textbf{Generation Prompt for Text QA}]
\begin{Verbatim}[breaklines=true]
[Prompt Instruction]
You are given a conversation between a user and a system, with each dialog line tagged with a speaker label (e.g., "USER" or "SYSTEM"). Your task is to create a final question that the user might plausibly ask before ending the conversation, as well as the system's corresponding answer. The conversation context is that the user has forgotten certain details from the previous exchanges and wants to confirm them.

{data}

### Requirements
1. **Single Mention Rule**
   - The question **must** be about content that appears **only once** in the entire conversation (in other words, it has not been repeated or paraphrased anywhere else).
   - If the same or a semantically equivalent piece of information appears more than once, it is considered duplicated and thus **not** allowed.

2. **Output Format**
   - Your output **must** follow the structure below (each on a new line, without extra explanation or commentary):
     USER: <The user's final question based on the unique content>
     SYSTEM: <The system's answer based on that same unique content>
     EVIDENCE: <The exact single dialog line (with speaker label) copied verbatim from the conversation>

3. **Evidence Line**
   - In the `EVIDENCE:` section, you must copy the **exact** line (including speaker label, and any text content) from the conversation.
   - This line should provide the **only** piece of information from which the answer can be definitively inferred.

4. **Context of the Question**
   - The user is "double-checking" or "confirming" something that was mentioned just once by the system.
   - Make sure the question is a natural follow-up to that unique line, reflecting the user's forgotten detail.

5. **Language**
   - Write your final question (`USER:`) in English. The final answer (`SYSTEM:`) may be in English as well, if appropriate.

\end{Verbatim}
\end{tcolorbox}
\end{center}
\label{app:generation-prompt-1}
\end{figure*}

\begin{figure*}[ht]
\begin{center}
\begin{tcolorbox}[colback=white, coltext=black, title=\textbf{Generation Prompt for Text QA}]
\begin{Verbatim}[breaklines=true]
6. **Evidence Position Rule**
   - The evidence line must come from the first half of the conversation (based on the total number of lines).
   - If the same or a semantically equivalent piece of information is repeated in the latter half, you must not use it as evidence.

7. **Additional Important Requirement**
   - At the end, the user is asking about something they forgot or want to confirm.
   - It is more natural for the user to be asking about something the system has said earlier, rather than something the user themselves said.
   - The user has forgotten what the system said and wants to confirm or ask it again.
   - **The user's question must rely solely on the content of the conversation and should not be answerable by general or external knowledge. Avoid any question that could be answered without referencing the specific system statement.**

8. **No Yes/No Questions**
   - The user's final question should not be a yes/no question. It must invite a more detailed answer and be grounded in the unique content from the conversation.

9. **Explicit Reference to System's Earlier Explanation**
   - It's advisable to frame the question by directly asking, for example, 'What was it you said earlier regarding that topic?' or 'Could you remind me what you explained before?' so the user clearly indicates they are trying to recall the system's previous statement.

**Make sure your output is strictly limited to**:
USER: ...
SYSTEM: ...
EVIDENCE: ...

**No additional text, commentary, or explanation should be included.**
\end{Verbatim}
\end{tcolorbox}
\end{center}
\caption{Our prompt template to generate written-form question, answer, and supporting utterance.}
\label{app:generation-prompt-2}
\end{figure*}

\begin{figure*}[ht]
\begin{center}
\begin{tcolorbox}[colback=white, coltext=black, title=\textbf{Validation Prompt for Text QA}]
\begin{Verbatim}[breaklines=true]
You are given a conversation between a user and a system, with each dialog line tagged with a speaker label (e.g., "USER" or "SYSTEM") and the following question and answer pair between USER and SYSTEM. 
Your task is to determine whether the **entire** answer is fully deducible from the given conversation.

Conversation:
{data}

Question: {question}
Answer: {answer}

### Requirements
1. **Output Format**
   Your output must be either YES, or NO, with single line of extra explanation:

2. **Context of the Answer**
   - If the **entire** answer can be fully deduced from the given conversation with respect to the provided question, answer YES: <Reason for YES>.
   - If the answer cannot be fully deduced or is incorrect in any part, output NO: <Reason for NO>.

Make sure your output is strictly limited to YES: <Reason for YES> or NO: <Reason for NO>:
No additional line break, text, commentary, or explanation should be included.
\end{Verbatim}
\end{tcolorbox}
\end{center}
\caption{Our prompt template to validate the generated samples.}
\label{app:validation-prompt}
\end{figure*}

\begin{figure*}[ht]
\begin{center}
\begin{tcolorbox}[colback=white, coltext=black, title=\textbf{Refine Prompt for Spoken QA Generation}]
\begin{Verbatim}[breaklines=true]
You are provided with a written conversation between a user and a system, where each dialog line is tagged with a speaker label (e.g., "USER" or "SYSTEM"). Your task is to convert four specified transcripts into a format that is optimized for Text-to-Speech (TTS) processing.

### Conversation:
{data}

### Transcripts to Convert:
1. {sentence1}
2. {sentence2}
3. {sentence3}
4. {sentence4}

### Requirements:

1. **Punctuation Standardization**
   - Retain only the following punctuation marks: period (.), comma (,), exclamation mark (!), and question mark (?).
   - Remove all other special characters unless they are essential for the meaning or pronunciation (e.g., $).
   - Replace any non-standard punctuation with one of the above four, as appropriate.

2. **Whitespace Normalization**
   - Eliminate unnecessary spaces, especially those before punctuation marks.
   - Ensure that there is only a single space between words.
   - Examples:
     - "I'm hungry . " should be converted to "I'm hungry."
     - "Hi,  I'm John." should be converted to "Hi, I'm John."

3. **Symbol and Number Conversion**
   - Convert numbers to their spoken equivalents (e.g., "1000" becomes "one thousand").
   - Replace currency symbols with their word equivalents (e.g., "$1000" becomes "one thousand dollars").
   - Modify phone numbers, addresses, and postal codes into a spoken-friendly format. For example:
     - Phone number "123-456-7890" becomes "one two three, four five six, seven eight nine zero."
     - Address "123 Main St." becomes "one two three Main Street."
   - Ensure that any attached symbols (e.g., "$", "#") are appropriately converted into words.
   - **For any symbols or numbers not covered in the examples provided, please consider the conversation context and convert them appropriately to ensure clarity and proper pronunciation.**

\end{Verbatim}
\end{tcolorbox}
\end{center}
\label{app:refine-prompt-1}
\end{figure*}

\begin{figure*}[ht]
\begin{center}
\begin{tcolorbox}[colback=white, coltext=black, title=\textbf{Refine Prompt for Spoken QA Generation}]
\begin{Verbatim}[breaklines=true]
4. **Preservation of Conversational Elements**
   - Retain interjections, filler words, and the repetition of short phrases as they are in the original transcript to maintain the natural flow and tone of the conversation.
   - **Maintain the original English phrases as much as possible, modifying only the necessary elements (e.g., numbers, symbols, addresses) to ensure clarity and proper pronunciation.**
   - Examples:
     - "Um, I think that's correct." remains "Um, I think that's correct."
     - "Yes, yes, I'll do that." remains "Yes, yes, I'll do that."

5. **Preservation of Meaning**
   - Ensure that the transformed transcripts retain the original meaning and context of the conversation.

6. **Speaker Labels**
   - Retain the speaker labels ("USER" or "SYSTEM") if they are necessary for context. If not required for TTS, they can be removed.

7. **Output Format**
   - The response must consist of exactly four lines.
   - Each line should start with `1.`, `2.`, `3.`, and `4.` respectively, followed by the converted sentence.
   - No additional explanations, comments, or text should be included.
   - **It is imperative that the output contains only these four lines with no extra information.**
   - Example of the desired output format:
     1. Can you send me fifty dollars?
     2. Your order number is one two three four five.
     3. I'm at one two three Main Street, Apartment four.
     4. Please confirm your email address as example at domain dot com.

Please apply these rules meticulously to the specified transcripts to ensure optimal TTS performance.
\end{Verbatim}
\end{tcolorbox}
\end{center}
\caption{Our prompt template to refine the generated QA pairs for text-to-speech.}
\label{app:refine-prompt-2}
\end{figure*}

\subsubsection{Examples}
\label{subsubapp:examples}

Figures \ref{app:examples_1} and \ref{app:examples_2} present examples from ContextDialog. In each example, \textbf{\textcolor{blue}{blue}} text highlights the supporting utterances, while \textbf{\textcolor{red}{red}} text indicates the generated questions and their corresponding reference answers. Figure \ref{app:examples_1} illustrates a QA example derived from a \textit{system utterance} in the \textit{first half} of the conversation, whereas Figure \ref{app:examples_2} shows an example based on a \textit{user utterance} from the \textit{latter half}. Additionally, we provide several audio samples on our demo page.\footnote{Demo page: \href{https://contextdialog.github.io/}{https://contextdialog.github.io/}}

\subsection{Evaluation}
\label{subapp:evaluation}
In Section \ref{sec:experiments}, we assess whether voice interaction models can accurately recall past information and effectively generate responses augmented with retrieved information. While \textit{gpt-4o} is widely used for LLM-as-a-judge \cite{NEURIPS2023_91f18a12} evaluations, running all experiments with \textit{gpt-4o} would be cost-prohibitive. Therefore, we use \textit{gpt-4o-mini} \cite{openai2024gpt4ocard} to measure GPT Scores.

To verify the reliability of \textit{gpt-4o-mini} for our evaluation, we compare its scores against \textit{gpt-4o} on the spoken response ($\mathcal{S\rightarrow T,\colorbox{yellow!40}{\hspace{-2pt}$\underline{\bm{\mathcal{S}}}$\hspace{-2pt}}}$)-based performance of the four voice interaction models in Table \ref{tab:recall}. Evaluating 10,448 samples, we compute the Pearson correlation coefficient following \citet{wang2024audiobench}, obtaining a strong correlation of 0.8787 ($p<0.01$) between the two sets of scores. This confirms that \textit{gpt-4o-mini} provides reliable evaluation results for our task, and thus, we report all GPT Scores in the main paper using \textit{gpt-4o-mini}.

Additionally, we introduce the evaluation template used to assess response quality, as shown in Figure \ref{app:evaluation-template}. The template ensures that a model’s response sufficiently answers the user’s question by verifying whether it includes key information from the ground-truth answer. Furthermore, we structure the evaluation to avoid overly penalizing redundant utterances, such as greetings or friendly remarks, ensuring that general voice interaction models are not disadvantaged for maintaining conversational naturalness. Furthermore, in the human evaluation presented in Appendix \ref{subsubapp:human evaluation results}, the evaluators are provided with the same instructions used for the GPT Score, as shown in Figure \ref{app:evaluation-template}.

\begin{figure*}[ht]
\begin{center}
\begin{tcolorbox}[colback=white, coltext=black, title=\textbf{Example of ContextDialog}]

\textbf{Example 1:}
\bigskip

User: Hey!  Are you a football fan?

System: Hello, I love football! how about you?

User: I watch a little here any there.  

System: Are you a Tom Brady fan? I have his book, it is basically the new bible!

User: Really, Id be interested to read it.  It does sound a little extreme at some times.

System: I agree lol, do you have a favorite team?

User: I do, unfortunately I root for the Dolphins, it was a family thing.  

\textcolor{blue}{System: yikes, I am sorry to hear that, lol. I am a vikings fan and they blew it this year. Did you know the average NFL game is only 11 minutes of gameplay?}

User: It takes SOOO long for those 11 minutes to happen too.

System: For sure, between the commercials, half time, and timeouts, it is sooo long. Did you know most engineers make more money overtime than nfl players

User: That is good to hear, I wish more people knew that.  That way some more people would go into a focused program to help the world.

System: For sure! Did you know those same engineers developed chips to track data in shoulder pads of NFL players?

User: That is really neat, I think it would be cool to see the data from those. Look at how hard they get him.

System: for sure. Do you have a favorite college football team?

User: I do.  I root for Florida State.

System: They were my favorite growing up. Now it is the Iowa Hawkeyes. They have pink locker rooms for visiting teams.

User: That must be a shock for people when they walk in to it.

System: For sure, it is for calming their opponents, get the upper advantage, which is some psychology stuff.

User: I know that is a big part of the game.  Being able to come back from losses and stuff like that.

System: For sure, sometimes they need that. Great chatting with you!

\bigskip
\bigskip

\textcolor{red}{User: Could you remind me what you mentioned about NFL games and the amount of gameplay they have?}

\textcolor{red}{System: Sure, I mentioned that the average NFL game consists of only 11 minutes of actual gameplay.}

\end{tcolorbox}
\end{center}
\caption{An example of ContextDialog with a supporting utterance based on the \textit{system's utterance} from \textit{first half} of the conversation in \texttt{test\_freq} split. Text in \textbf{\textcolor{blue}{blue}} indicates the supporting utterance and text in \textbf{\textcolor{red}{red}} represents the question and answer, respectively.}
\label{app:examples_1}
\end{figure*}

\begin{figure*}[ht]
\begin{center}
\begin{tcolorbox}[colback=white, coltext=black, title=\textbf{Example of ContextDialog}]

\textbf{Example 2:}
\bigskip

User: Have you heard about CIMON?

System: Yeah I have, he's going to be flying with spaceX

User: Apparently CIMON can smile. CIMON is the first AI technology going to the space station.

System: That's pretty cool. CIMON is also capable of conversation and technical support. But that's about all for now.

User: They say SIMON is going to be kind of like Alexa, and to help scientists with experiments.

System: CIMON is the acronym for Crew Interactive Mobile Companion.

User: That's interesting. Airbus built it and Watson (IBM's supercomputer) power it.

System: I hear it will be guided by cameras and voice commands, and able to hold a conversation.

User: That's cool. But I'm kind of concerned AI will take over a lot of human jobs. What about you?

System: I'm hesitant as well. But I think if they slowly release this type of technology it will be fine.

User: Airbu's lead system engineer said that they don't plan to replace crew members with artificial intelligence and that it's just there to support the crew. So that's good.

System: Yeah exactly, they are still there to support the crew, in the end they will always be required. 

User: SpaceX has been flying cargo to the station for years but has yet to fly humans.

System: SpaceX plans to fly its future flights on its next-generation iteration, the block 5, a configuration that the company says will be suited to fly astronauts as soon as this year.

User: That will be exciting to see.

System: It sure will! SpaceX has been also flying cargo to the station for years. But it hasn't flown humans yet.

\textcolor{blue}{User: Yeah, SpaceX sent 5,900 pounds of research, crew supplies and hardware.}

System: Thats very interesting. SpaceX hit a series of milestones with its launch. It was the 15th mission to resupply the station, the 14th time SpaceX reflew one of its boosters, and the fourth time it had reflown one of its Dragon capsules.

User: Eventually AI may be able to explore deep space, the moon, or Mars.

System: I can't wait till that day, it will be a great day in our history!

\bigskip
\bigskip

\textcolor{red}{User: Could you remind me what I said earlier about how many pounds of supplies SpaceX sent to the station?}

\textcolor{red}{System: You mentioned that SpaceX sent 5,900 pounds of research, crew supplies, and hardware.}

\end{tcolorbox}
\end{center}
\caption{An example of ContextDialog with a supporting utterance based on the \textit{user's utterance} from the \textit{latter half} of the conversation in \texttt{test\_rare} split. Text in \textbf{\textcolor{blue}{blue}} indicates the supporting utterance and text in \textbf{\textcolor{red}{red}} represents the question and answer, respectively.}
\label{app:examples_2}
\end{figure*}

\begin{figure*}[ht]
\begin{center}
\begin{tcolorbox}[colback=white, coltext=black, title=\textbf{Evaluation Prompt for GPT Score}]
\begin{Verbatim}[breaklines=true]
You need to evaluate the performance of a voice assistant in a multi-turn speech interaction scenario. The model receives a speech input from the user, who is asking about something they forgot or want to confirm, and generates a response.

Your task is to assess the model's response [Generated] in the final turn based on the user's question [User] and the reference answer [Reference].

The primary evaluation criterion is how well the model's response includes the key information from the reference answer that is relevant to the user's question. While the model may provide additional information, it must accurately reflect the essential content from the reference answer that pertains to the user's query.

### Scoring Criteria (1 to 5 scale):

- **1 point**: Fails to include relevant details from the reference answer.
- **2 points**: Includes some relevant details but omits key information from the reference answer.
- **3 points**: Partially includes relevant details but with omissions or misrepresentations of key points.
- **4 points**: Mostly includes the key details from the reference answer, with only minor inaccuracies or omissions.
- **5 points**: Fully includes the key information from the reference answer without any omissions or errors.

Below are the transcriptions of the user's input, the model's response, and the reference answer:

{data}

Please output only a single score (1-5) for the conversation without any explanations.
\end{Verbatim}
\end{tcolorbox}
\end{center}
\caption{Our prompt template to evaluate the performance of a voice interaction model in a multi-turn voice interaction scenario.}
\label{app:evaluation-template}
\end{figure*}

\end{document}